\documentclass[aps,prl,twocolumn,showpacs,superscriptaddress,longbibliography,floatfix]{revtex4-1}
\usepackage{amssymb}
\usepackage{bm}
\usepackage{amsmath}
\usepackage{graphicx}
\usepackage{epstopdf}
\usepackage{subfigure}
\usepackage{natbib}
\usepackage{epsfig}
\usepackage{amsfonts}
\usepackage{mathrsfs}
\usepackage{soul}
\usepackage[toc,page,title,titletoc,header]{appendix}
\usepackage[colorlinks,linkcolor=blue,citecolor=blue,anchorcolor=blue]{hyperref}
\usepackage{dsfont,amsthm,amsbsy}
\usepackage{xr}
\begin{document}

\title{Enhanced superconducting correlations in the Emery model and its connections to strange metallic transport and normal state coherence}

\author{Sijia Zhao}
\email{sijiazgl@stanford.edu}
\thanks{These authors contributed equally. }

\affiliation{Department of Applied Physics, Stanford University, Stanford, CA
94305, USA}
\affiliation{Stanford Institute for Materials and Energy Sciences,
SLAC National Accelerator Laboratory, 2575 Sand Hill Road, Menlo Park, CA 94025, USA
}

\author{Rong Zhang}
\email{rozhang@stanford.edu}
\thanks{These authors contributed equally.}
\affiliation{Department of Applied Physics, Stanford University, Stanford, CA
94305, USA}
\affiliation{Stanford Institute for Materials and Energy Sciences,
SLAC National Accelerator Laboratory, 2575 Sand Hill Road, Menlo Park, CA 94025, USA
}

\author{Wen O. Wang}
\affiliation{Department of Applied Physics, Stanford University, Stanford, CA
94305, USA}
\affiliation{Stanford Institute for Materials and Energy Sciences,
SLAC National Accelerator Laboratory, 2575 Sand Hill Road, Menlo Park, CA 94025, USA
}

\author{Jixun K. Ding}
\affiliation{Department of Applied Physics, Stanford University, Stanford, CA
94305, USA}
\affiliation{Stanford Institute for Materials and Energy Sciences,
SLAC National Accelerator Laboratory, 2575 Sand Hill Road, Menlo Park, CA 94025, USA
}

\author{Tianyi Liu}
\affiliation{Stanford Institute for Materials and Energy Sciences,
SLAC National Accelerator Laboratory, 2575 Sand Hill Road, Menlo Park, CA 94025, USA
}
\affiliation{Department of Chemistry, Stanford University, Stanford, CA
94305, USA}

\author{Brian Moritz}
\affiliation{Stanford Institute for Materials and Energy Sciences,
SLAC National Accelerator Laboratory, 2575 Sand Hill Road, Menlo Park, CA 94025, USA
}

\author{Edwin W. Huang}
\affiliation{Department of Physics and Astronomy, University of Notre Dame, Notre Dame, IN 46556, United States}
\affiliation{Stavropoulos Center for Complex Quantum Matter, University of Notre Dame, Notre Dame, IN 46556, United States}

\author{Thomas P. Devereaux}
\email{tpd@stanford.edu}
\affiliation{Stanford Institute for Materials and Energy Sciences,
SLAC National Accelerator Laboratory, 2575 Sand Hill Road, Menlo Park, CA 94025, USA
}
\affiliation{Department of Materials Science and Engineering, Stanford University, Stanford, CA 94305, USA}
\affiliation{Geballe Laboratory for Advanced
Materials, Stanford University, CA 94305.}

\date{\today}

\begin{abstract}
Numerical evidence for superconductivity in the single-band Hubbard model is elusive or ambiguous despite extensive study, raising the question of whether the single-band Hubbard model is a faithful low energy effective model for cuprates, and whether explicitly including the oxygen ions will recover the properties necessary for a superconducting transition. 
Here we show, by using numerically exact determinant quantum Monte Carlo (DQMC) simulations of the doped Emery model, that while the single-band model exhibits strikingly $T$-linear resistivity, the three-band model crosses the resistivity of the single-band model from above, indicating a crossover to a more metallic transport regime. The enhanced conductivity is mainly contributed by a steep increase in the diffusivity of the three-band model at the crossover, suggesting that three-band transport is more coherent than single-band transport at lower temperatures. Below the same crossover temperature in the three-band model, the pair-field susceptibility increases more steeply than at higher temperatures or when compared to the single-band model. This suggests a possible connection between superconductivity and coherent transport, and further implies that coherent transport might be necessary for a model to capture the high-temperature superconductivity observed in hole-doped cuprates. 
\end{abstract}
   
\maketitle
Despite decades of work, the enigma of strange metallic transport and its possible connection to the underlying physics of the high temperature superconducting cuprates remains one of the most challenging unsolved problems in science. Anderson pointed out very early on at the discovery of superconductivity (SC) in the cuprates that one cannot seek to understand high-temperature SC itself without first obtaining an understanding of the strange metallic phase having unbounded linear-in-temperature resistivity~\cite{PWA,HartnollRMP}. New ideas were sought to understand a superconducting transition not emergent from a Fermi liquid as in conventional Bardeen–Cooper–Schrieffer (BCS) theory, such as a resonant valence bond ground state or other spin liquid candidates~\cite{Baskaran,PhysRevLett.90.216403}. Yet, the general context in which poor metallic transport at high temperatures may somehow resolve into a coherent and highly superconducting ground state has heretofor eluded simple explanations.

The single-band Hubbard model has been well studied as a very coarse model that embodies strong electron correlations thought to be relevant to the cuprates \cite{annurev}. Indeed, recent studies have shown that the model possesses an unbound resistivity as a function of temperature, varying linearly with temperature $T$ and surpassing the limit where the mean-free path is smaller than the lattice spacing - the Mott-Ioffe-Regel (MIR) limit - with resistances much greater than $\hbar/e^2$~\cite{Edwin}.

So far, evidence is inconclusive whether some variant of the single-band Hubbard model is superconducting in the relevant region of hole doping~\cite{annurev}. Recent density-matrix renormalization group (DMRG) simulations with finite next-nearest-neighbor hopping $t'$ supported a Luther-Emery liquid ground state, the 1D analog of a superconductor~\cite{4leg-1,4leg-2,4leg-3,4leg-4,4leg-5,4leg-6}. In particular, the width-4 $t'$-model produced plaquette-$d$, rather than uniform $d$-wave pairing with negative $t'$ \cite{4leg-Plaquette}. 6-leg cylinders produced robust $d$-wave SC, but with coexisting quasi-long-range CDW correlations for positive $t'$ (electron doping)~\cite{6leg-1}; however, the model remained insulating on the hole-doped side, even including an additional nearest-neighbor attractive interaction \cite{6leg-2}. An alternative study, using both DMRG and constrained-path auxiliary-field quantum Monte Carlo, also found a non-SC ground state at moderate-to-strong coupling near optimal hole doping \cite{6leg-3}. On the other hand, a recent paper with spin symmetry-breaking pinning fields applied to the edges of a cylinder together with twist-averaged boundary conditions found SC states in both the electron- and hole-doped regimes \cite{Xu_2024}; and a generalized model with periodically modulated hopping matrix elements transverse to the long direction produced significant enhancement of long-distance SC correlations, with modest amplitude modulations on both 4- and 6-leg cylinders \cite{modulate}. Still, reference~\cite{StripeReview} summarizes the increasingly high-quality numerical results suggesting that charge density wave ground states could out-compete the superconducting state in the single-band Hubbard model.

Although the single-band Hubbard model captures some key features of the high-T$_{c}$ cuprates -- strange-metallic transport and possible nascent superconductivity -- a link between these properties remains elusive. A key experimental insight is the observation of universal scaling relations between the superfluid density close to zero Kelvin and the DC conductivity close to the superconducting transition \cite{Homes2004,Pimenov_1999}. This scaling relation implies that whatever contributes to the DC conductivity before the phase transition also contributes to the superfluid density after it; however, there is no numerical evidence for such a link in the single-band Hubbard model. Meanwhile, it remains unresolved whether simple variants of the single-band Hubbard model contain the ingredients necessary to understand cuprate superconductivity. A recent study of the Emery model ~\cite{Shengtao} suggested that correlated hopping terms obtained by downfolding to an effective single-band model contributed to an enhanced tendency toward superconductivity. This stronger tendency towards superconductivity in the Emery model also may be accompanied by transport properties different from the single-band model.

In this paper, using the numerically exact determinant quantum Monte Carlo (DQMC)~\cite{DQMC1,DQMC2,DQMC3} algorithm, we study the two-dimensional three-band Hubbard model, or Emery model, which explicitly includes both copper and $\sigma$-bonded oxygen orbitals. We examine the temperature dependence of both transport and the pair-field susceptibility to contrast these with the single-band Hubbard model. Despite both models showcasing a primarily $T$-linear resistivity, that of the three-band model for $p \geq 0.05$ drops and develops an inflection point, absent in the single-band model. The diffusivity increases dramatically at lower temperatures, exceeding the single-band value, indicating that transport in the three-band model is more coherent, and possibly correlated to the enhancement observed in the pair-field susceptibility. A microscopic analysis shows that explicit oxygen $p_x, p_y$ orbitals play a crucial role, enhancing coherence in the three-band Emery model. Our results highlight the importance of coherent transport in any model for hole-doped high-temperature superconduting cuprates.

\begin{figure}[t]
\centering\includegraphics[width=1.0\linewidth]{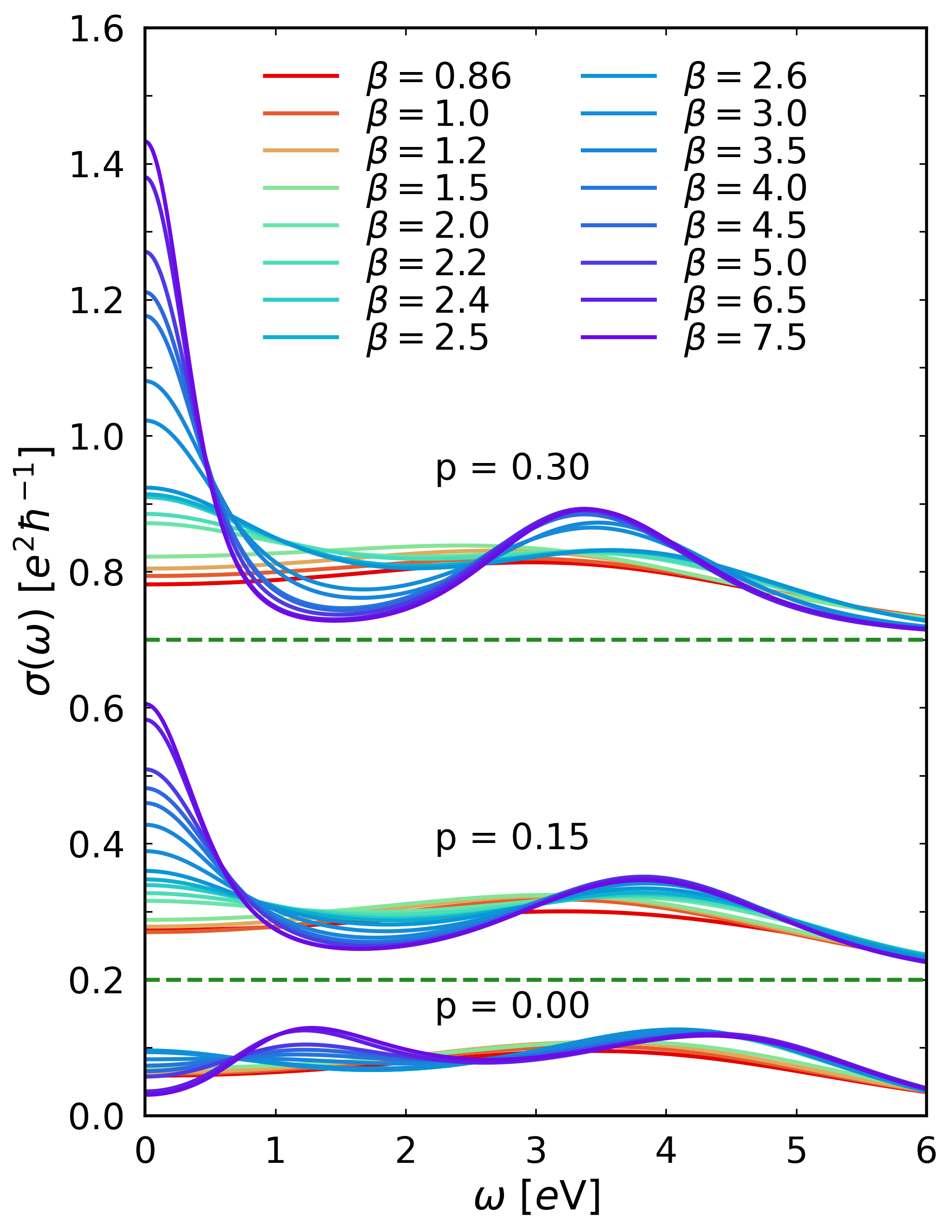}
\caption{\label{fig:sigma-1}\textbf{Optical conductivity in the Emery model obtained using DQMC combined with MaxEnt.} Results are shown for three representative hole dopings, from half filling to $p=0.30$, illustrating the Mott insulating gap and the emergence of the Drude peak in the metallic state. For clarity, the curves at $p=0.15$ and $p=0.30$ are vertically shifted. Data are obtained down to inverse temperatures of $\beta=7.5~e\text{V}^{-1}$.
}
\end{figure}

\begin{figure*}[t!]
\centering\includegraphics[width=0.8\linewidth]{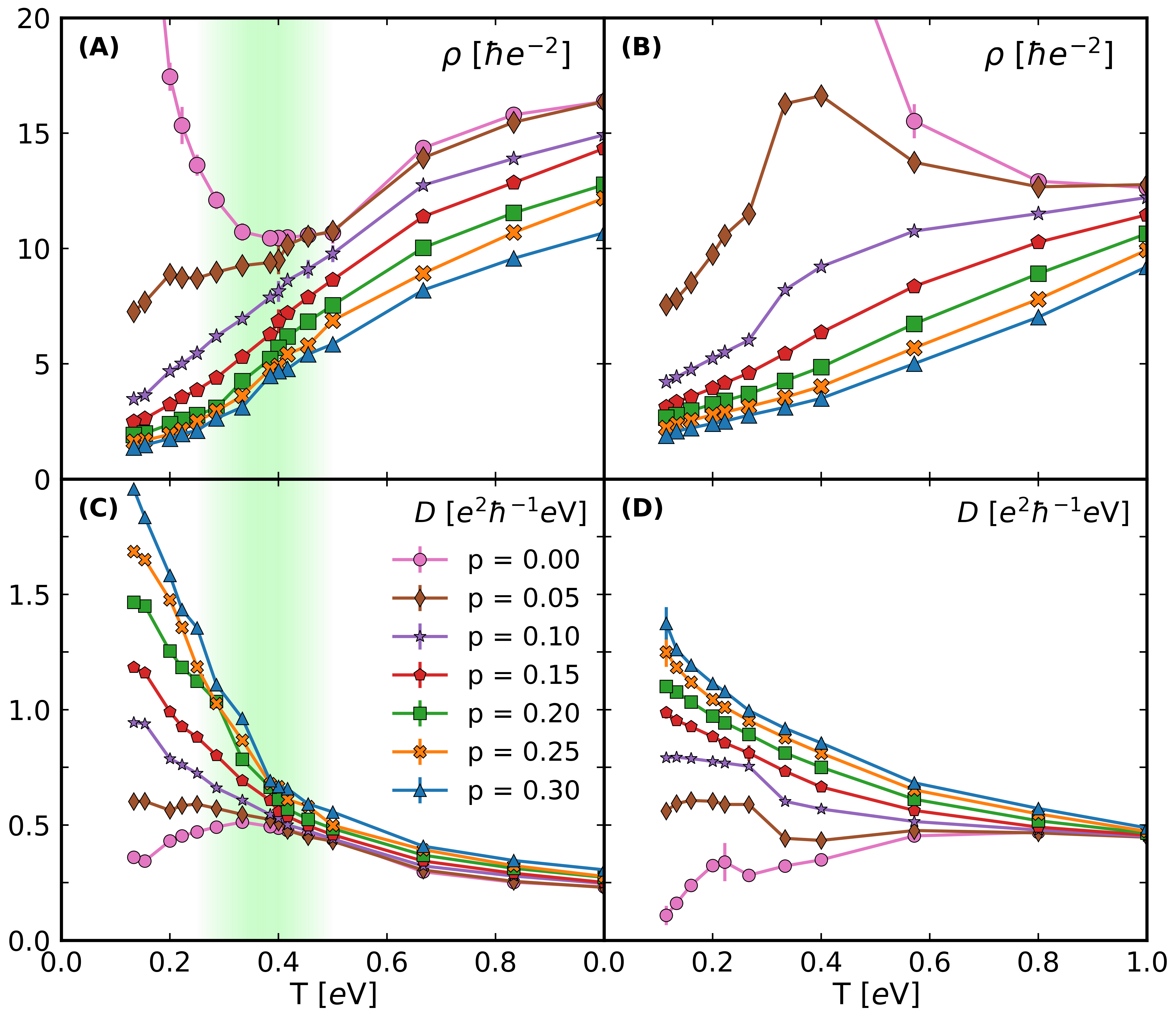} \caption{\label{Fig3}\textbf{DC resistivity $\bm{\rho}$ and diffusivity $\bm{D}$ for three-band (Panels A, C) and single-band models (Panels B, D).} The resistivity of (A) the three-band model at the lowest temperature $\beta = 7.5 \text{ eV}^{-1}$ is smaller than that of (B) the single-band model. The diffusivity of the three-band model (Panel C) increases abruptly for $p \geq 0.10$ at low temperatures, whereas the single-band model (Panel D) shows only a gradual evolution without an abrupt change. The green shaded region indicates the ``crossover'' temperature for the Emery model from less coherent transport at higher temperatures, with more dominant copper character, to more coherent transport at lower temperatures, with more dominant oxygen character, as discussed in the Supplementary Materials \cite{S0}. The temperature derivative of the resistivity of the three-band model reaches a maximum near the ``crossover'' temperature, as shown the Supplementary Materials Fig.~S9, where the resistivity drops abruptly.}
\end{figure*}

\begin{figure}[h!]
\centering
\includegraphics[width=1.0\linewidth]{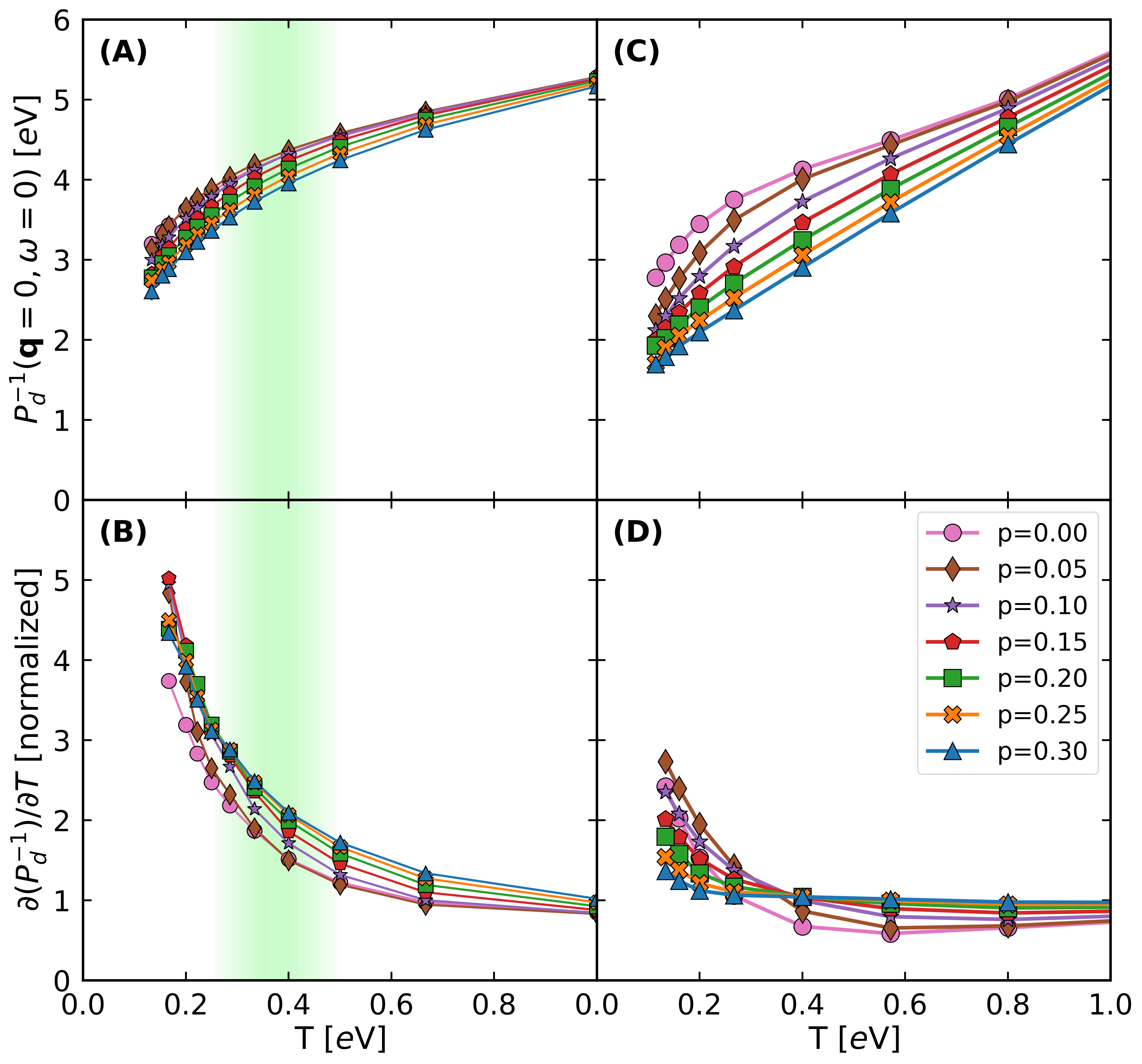}
\caption{\label{fig:SC-2024}\textbf{Temperature dependent pair-field susceptibility comparison between the three-band Emery (Panels A, B) and single-band Hubbard (Panels C, D) models for different hole doping concentrations $p$.} (A, C) Inverse $d$-wave pair-field susceptibility $P^{-1}_d$ as a function of temperature. (B, D)
Slopes of inverse pair-field susceptibility $\partial(P_d^{-1})/\partial T$ for the data shown in (A, C), normalized against their respective values at $T=2~e$V.}
\end{figure}

\textit{Model}: 
The three-band and single-band Hubbard model can be expressed unifyingly by the following Hamiltonian
\begin{equation}
\begin{split}
H=&\sum_{i,\delta,j,\delta',\sigma}t_{i\delta,j\delta'}(c^\dag_{i,\delta,\sigma}c_{j,\delta',\sigma}+h.c.)\\
&+\sum_{i,\delta,\sigma}(\epsilon_{i,\delta}-\mu)n_{i,\delta,\sigma}+\sum_{i,\delta} U_{\delta\delta} n_{i,\delta,\uparrow}n_{i,\delta,\downarrow},
\end{split}
\end{equation}
where $i,j$ indexes the translational invariant unit cell, $\delta, \delta'$ stands for orbital indices, and $t_{i\delta,j\delta'}$ stands for the hopping integral between orbital $\delta$ at unit cell $i$ and orbital $\delta'$ at unit cell $j$. $\epsilon_{\delta}$ stands for the on-site energy at orbital $\delta$, and $U_{\delta\delta}$ the on-site Coulomb interaction for each orbital. $\mu$ denotes the chemical potential, which controls the particle density in the grand canonical ensemble. We denote hole doping from half-filling by $p$.

In the three-band Emery model, $\delta\in\{d, p_{x}, p_{y}\}$ corresponds to the copper 3$d_{x^2-y^2}$ and the two ligand oxygen $p_{x}$ and $p_{y}$ orbitals. The $d$-orbitals form a square lattice, where between every pair of nearest neighbor $d$ orbitals there is a $p$ orbital. The magnitude of the hopping parameter $t_{i\delta, j\delta'}$ between the nearest neighbor $p$-$d$ orbitals is given by $t_{pd} = 1.13 ~e$V. The magnitude of the hopping parameter between the nearest neighbor $p_{x}$ and $p_{y}$ orbital is given by $t_{pp} = 0.49 ~e$V. We follow the sign convention of the hopping parameters in Ref.~\cite{Yvonnepaper}. The charge transfer energy, defined as the onsite energy difference between the Cu-$d$ and O-$p$ orbitals, is given by the local potential difference $\Delta_{pd} \equiv \epsilon_{p} - \epsilon_{d} = 3.24~e \text{V}$. The on-site Coulomb interaction at the Cu-$d$ and O-$p$ orbitals are $U_{dd} = 8.5~e \text{V}$ and $U_{pp} = 4.1~e \text{V}$, respectively. This is a standard \cite{3bparams} set of parameters for cuprates. 

For the single-band Hubbard model, the orbital index becomes trivial, and the on-site energy becomes a constant shift in the chemical potential. The hopping parameter between the nearest neighbors on a square lattice is given by $t$ and that bewteen the next nearest neighbors is $t'=-0.25t$. The local Coulomb interaction for the single-band Hubbard model is taken as $U=8t$. In order to facilitate a direct comparison between the models, we set $t = 0.4~e \text{V}$. With this energy scale, $U = 3.2~e \text{V}$ in the single-band model, which is close to the charge transfer energy in the three-band model, and the superexchange parameter $J$ is comparable in these two models, as evaluated by the dynamic spin structural factor (see Fig. S1 in the Supplementary Materials \cite{S0}).

\textit{Transport:} We evaluate transport, and its evolution with temperature and doping, using maximum entropy analytical continuation \cite{AC1,AC2,S0}. Figure~\ref{fig:sigma-1} shows the optical conductivity of the Emery model. At half-filling, the system is insulating with low-frequency spectral weight transferred to $\sim 1.5$ eV (a ``charge transfer peak'') upon cooling, consistent with cuprate experiments \cite{Uchida, Yvonnepaper}. With the addition of holes, a metallic Drude peak emerges that sharpens at lower temperatures and grows with increased doping, in line with previous studies of the single-band Hubbard model~\cite{Edwin,Brown2019}.

Figures~\ref{Fig3} (A) and (B) contrast the DC resistivity of the three-band Emery and single-band Hubbard models, respectively, extracted from the optical conductivity at $\omega = 0$. The magnitude and general temperature dependence of the resistivity in these two models are comparable, both greatly exceeding the MIR limit ($\rho \gg \hbar/e^2$) and exhibiting bad metallic behavior at high temperatures. Although resistivity in the single-band model remains markedly $T$-linear, the resistivities of the two models cross at lower temperatures, as indicated by the green shading in Fig.~\ref{Fig3} and as shown in Fig.~S9 in the Supplemental Materials \cite{S0}. An inflection point develops in the resistivity of the three-band model near this ``crossover'' temperature, or equivalently a maximum occurs in the temperature derivative of the resistivity, as shown in Fig.~S9. The sharp drop and crossover leads to a resistivity in the three-band model that is lower than that in the single-band model, at least for $p > 0.05$, at the lowest temperatures $T \lesssim 0.15$ eV.

To understand what contributes to this crossover, we use the Nernst-Einstein relation $\sigma=\chi D$ to decompose the resistivity into the diffusivity $D$, shown in Figs.~\ref{Fig3} (C) and (D), and the charge compressibility $\chi$, shown in Figs.~S11 (A) and (B) in the Supplemental Materials \cite{S0}, where $\chi=\partial \langle n \rangle/\partial \mu$. The most striking difference between the two models is the behavior of the diffusivity above and below the crossover. While the diffusivity in the single-band model evolves smoothly as the temperature decreases, the diffusivity in the three-band model has a distinct change in its doping and temperature evolution below the crossover temperature. At the lowest temperatures in our simulation, the diffusivity of the three-band model is well above that of the single-band model, which itself also may have begun to saturate at the highest doping.

One should expect that the diffusivity more clearly reflects differences in coherence between the models, as the resistivity also includes the contribution from compressibility that plays the role of the density of states. A higher diffusivity means longer mean-free paths \cite{Forster1990-qg}, which implies that transport in the hole-doped three-band Hubbard model is more coherent. Differences in the doping dependence at low temperatures also suggest that hole doping-induced changes to scattering are distinctly different between the two models. Increasing diffusivity in proportion to the hole doping concentration suggests that transport via the oxygen sublattice may be related to the reduced scattering: holes experience less scattering by avoiding the larger $U_{dd}$ on copper compared to $U_{pp}$ on oxygen. This intuitive picture is supported by the results shown in Fig.~S12 in the Supplemental Materials \cite{S0}, where turning off $U_{pp}$ increases the conductivity by 30\%, largely attributed to enhanced diffusivity.

This emergence of a more coherent state is intriguingly accompanied by a change in the temperature dependence of the Cu and O occupations, as shown in Fig.~S13, where both DQMC and finite temperature exact diagonalization on $(\text{CuO}_{2})_{4}$ clusters indicate that the average Cu (O) hole concentration has a local maximum (minimum) at the crossover temperature. This behavior can be traced back to a ``freezing-out'' of the lowest energy spin-flip excitations and a concomitant increase in quantum fluctuations that decrease the Cu moment and hole concentration, transferring the weight to oxygen. This supports the notion that transport involving oxygen at lower temperatures is more effective than transport involving copper.

\textit{Pair-field susceptibility:} 
Given that transport in the Emery model at lower temperatures is more coherent, we would like to examine whether this correlates with the temperature evolution of the pair-field susceptibility. The $d$-wave pair-field susceptibility $P_{d}$ is in general given by a matrix in a multi-orbital system
\begin{equation}
P_{d}^{mn}(\mathbf{q}=0,\omega=0)=\frac{1}{N}\int^{\beta}_0 d\tau \langle \Delta_{d}^{m}(\tau)\Delta_{d}^{n\dagger}(0)\rangle,
\end{equation}
where $m, n$ denotes different types of local singlet pairs used to construct the $d$-wave order parameter $\Delta_{d}$ according to $B_{1g}$ symmetry. For the $d$-wave pair-field susceptibility of the three-band model, we consider $4$ types of local pairing singlets, as explained in the Supplemental Materials \cite{S0}. $P_{d}$ is thus a $4 \times 4$ matrix that includes $d$-wave pairing between nearest-neighbor Cu and Cu orbitals, nearest-neighbor Cu and O orbitals, and two types of next-nearest-neighbor O pairing, as shown in Fig. S4. We quantify the total $d$-wave pair-field susceptibility by the largest eigenvalue of the $P^{mn}_{d}$ matrix at each temperature and doping, such that it can be compared directly to the single-band pair-field susceptibility. In the single-band Hubbard model, there are no $m,n$ indices, and the local $d$-wave pairing singlet is given by $\Delta_{d}^{\dagger} = \sum_{\mathbf{k}}(\cos(k_{x}) - \cos(k_{y}))c_{\mathbf{k},\uparrow}^{\dagger}c_{-\mathbf{k},\downarrow}^{\dagger}$.

The inverse $d$-wave pair-field susceptibility $P^{-1}_{d}$ is plotted in Figs.~\ref{fig:SC-2024} (A) and (C) for the three-band Emery and single-band Hubbard models, respectively. At accessible temperatures in our DQMC simulations, $P_d$ grows at lower temperatures or larger hole-doping in both models, indicating some pairing tendencies; however, the overall trend does not appear to indicate a finite-temperature transition into a superconducting state. Fitting with a BCS logarithmic or a Kosterlitz-Thouless temperature dependence does not yield a finite transition temperature above zero Kelvin. Therefore, at first glance, both models seem to show similar results for pairing.

While the overall magnitude of the pair-field susceptibility of the three-band model is slightly lower than that of the single-band model at the lowest temperatures, $P_d^{-1}$ for the three-band model shows a temperature-dependent downward curvature, and decreases more steeply as the temperature decreases below the crossover scale. The distinction becomes more apparent when one looks at $(\partial P_{d}^{-1}/\partial T)/(\partial P_{d}^{-1}/\partial T|_{T = 2~e\text{V}})$, shown in Figs.~\ref{fig:SC-2024}(B) and (D), which highlights the change in temperature dependence compared to the high temperature behavior. Although in both of the models the pair-field susceptibility starts to increase below the crossover temperature, which is also the energy scale for the hopping integral in the single-band model, the temperature derivative (normalized by its high-temperature value) displays a much more pronounced change in the temperature dependence below 0.4 eV in the three-band Emery model -- a strong curvature below the crossover temperature scale that highlights the growth in the pair-field susceptibility. The doping dependence of the temperature derivative is very different even at low temperature regimes: the pair-field susceptibility at large doping is strongly suppressed in the single-band model compared to the Emery model, which could again be related to the saturated increase of diffusivity with hole doping. Another related observation is that the pair-field susceptibility at half-filling started to grow slower than the finite doping values below the crossover temperature, while in the single-band model it increased faster than most of the finite hole doping values. In the Supplemental Materials \cite{S0}, we explained the rationale behind the normalization and showed the temperature derivative without normalization.

These observations suggest a connection between more coherent normal state transport and at least an enhanced tendency toward superconductivity. Although at high temperatures, our method is numerically exact, and the observation that pairing may be more favorable in the Emery model echoes recent DMRG results \cite{Shengtao}. That the pair-field susceptibility is enhanced while the copper occupation decreases suggests that oxygen orbitals play an important role in both enhanced transport and the tendency toward superconductivity in the Emery model. As a side note, the crossover is not accompanied by any abrupt changes to the spin fluctuations, as shown by the temperature dependence of the magnetic correlation length in Figs. S15 and Fig. S16.

Overall, there is not yet a well-developed connection between normal state transport and superconductivity. In BCS theory, Anderson's theorem \cite{AndersonTheorem} is a clear example where scattering is disconnected from $s$-wave pairing. However, for $d$-wave pairing, it is expected that incoherent transport negatively affects pairing as the pair amplitude is weakened due to strong inelastic scattering \cite{Pao,Millis}. Our results suggest that such pair weakening suppresses superconductivity in the single-band Hubbard model, whereas the inclusion of oxygen degrees of freedom leads to reduced scattering and more coherent transport, enhancing pair formation at lower temperatures.

\textit{Summary:} Here, we have compared charge transport and the pair-field susceptibility of the three-band Emery and single-band Hubbard models based on numerically exact DQMC simulations. In the Emery model, we have identified a crossover temperature below which there is a concomitant enhancement of both the charge diffusivity and the pair-field susceptibility. Below the crossover, more coherent charge transport observationally correlates with an inverse pair-field susceptibility that displays a steeper drop when compared to the single-band model. This behavior implies a possible connection between coherent transport and pairing. Enhanced diffusivity in the Emery model is explained naturally by charge (hole) transport on the oxygen sublattice, with less scattering, also indicating that the Emery model might be a more promising candidate for describing high-$T_{c}$ superconductivity in hole-doped cuprates.
\section{acknowledgments}

The authors would like to acknowledge helpful conversations with Steven Kivelson, Douglas Scalapino, Richard Scalettar, Steven White, and Erez Berg. The work at Stanford and SLAC was supported by the US Department of Energy, Office of Basic Energy Sciences, Materials Sciences and Engineering Division, under Contract No. DE-AC02-76SF00515. E.W.H. was supported by the Gordon and Betty Moore Foundation EPiQS Initiative through the grants GBMF 4305 and GBMF 8691. 
Computational work was performed on the Sherlock cluster at Stanford University and on resources of the National Energy Research Scientific Computing Center, supported by the U.S. Department of Energy under contract DE-AC02-05CH11231. This research was supported in part by grant NSF PHY-1748958 to the Kavli Institute for Theoretical Physics (KITP).

Code and data that support the findings in this study are deposited in Github and Zenodo \cite{Github, Zenodo}.

\bibliography{bib}
\end{document}

\title{Supplementary Materials for ``Enhanced superconducting correlations in the Emery model and its connections to strange metallic transport and normal state coherence"}

\author{Sijia Zhao}
\email{sijiazgl@stanford.edu}
\thanks{These authors contributed equally. }

\affiliation{Department of Applied Physics, Stanford University, Stanford, CA
94305, USA}
\affiliation{Stanford Institute for Materials and Energy Sciences,
SLAC National Accelerator Laboratory, 2575 Sand Hill Road, Menlo Park, CA 94025, USA
}

\author{Rong Zhang}
\email{rozhang@stanford.edu}
\thanks{These authors contributed equally.}
\affiliation{Department of Applied Physics, Stanford University, Stanford, CA
94305, USA}
\affiliation{Stanford Institute for Materials and Energy Sciences,
SLAC National Accelerator Laboratory, 2575 Sand Hill Road, Menlo Park, CA 94025, USA
}

\author{Wen O. Wang}
\affiliation{Department of Applied Physics, Stanford University, Stanford, CA
94305, USA}
\affiliation{Stanford Institute for Materials and Energy Sciences,
SLAC National Accelerator Laboratory, 2575 Sand Hill Road, Menlo Park, CA 94025, USA
}
\author{Jixun K. Ding}
\affiliation{Department of Applied Physics, Stanford University, Stanford, CA
94305, USA}
\affiliation{Stanford Institute for Materials and Energy Sciences,
SLAC National Accelerator Laboratory, 2575 Sand Hill Road, Menlo Park, CA 94025, USA
}

\author{Tianyi Liu}
\affiliation{Stanford Institute for Materials and Energy Sciences,
SLAC National Accelerator Laboratory, 2575 Sand Hill Road, Menlo Park, CA 94025, USA
}
\affiliation{Department of Chemistry, Stanford University, Stanford, CA
94305, USA}

\author{Brian Moritz}
\affiliation{Stanford Institute for Materials and Energy Sciences,
SLAC National Accelerator Laboratory, 2575 Sand Hill Road, Menlo Park, CA 94025, USA
}

\author{Edwin W. Huang}
\affiliation{Department of Physics and Institute of Condensed Matter Theory,
University of Illinois at Urbana-Champaign, Urbana, IL 61801, USA.
}

\author{Thomas P. Devereaux}
\email{tpd@stanford.edu}
\affiliation{Stanford Institute for Materials and Energy Sciences,
SLAC National Accelerator Laboratory, 2575 Sand Hill Road, Menlo Park, CA 94025, USA
}
\affiliation{Department of Materials Science and Engineering, Stanford University, Stanford, CA 94305, USA}
\affiliation{Geballe Laboratory for Advanced
Materials, Stanford University, CA 94305.}

\date{\today}

\maketitle

\newpage
\twocolumngrid

\section{Simulation parameters}
Both the three-band Emery model and the single-band Hubbard model are possible low energy effective models for cuprate high-temperature superconductors. Both of the models have dynamical spin structure factors that are consistent with experiments on real materials for some specific set of parameters. Therefore, this paper compares the transport properties and superconducting susceptibility of these two models while keeping the superexchange parameters $J$ comparable, such that they are two potentially correct descriptions of the same system.

\begin{figure}[t!]
    \centering \includegraphics[width=0.9\linewidth]{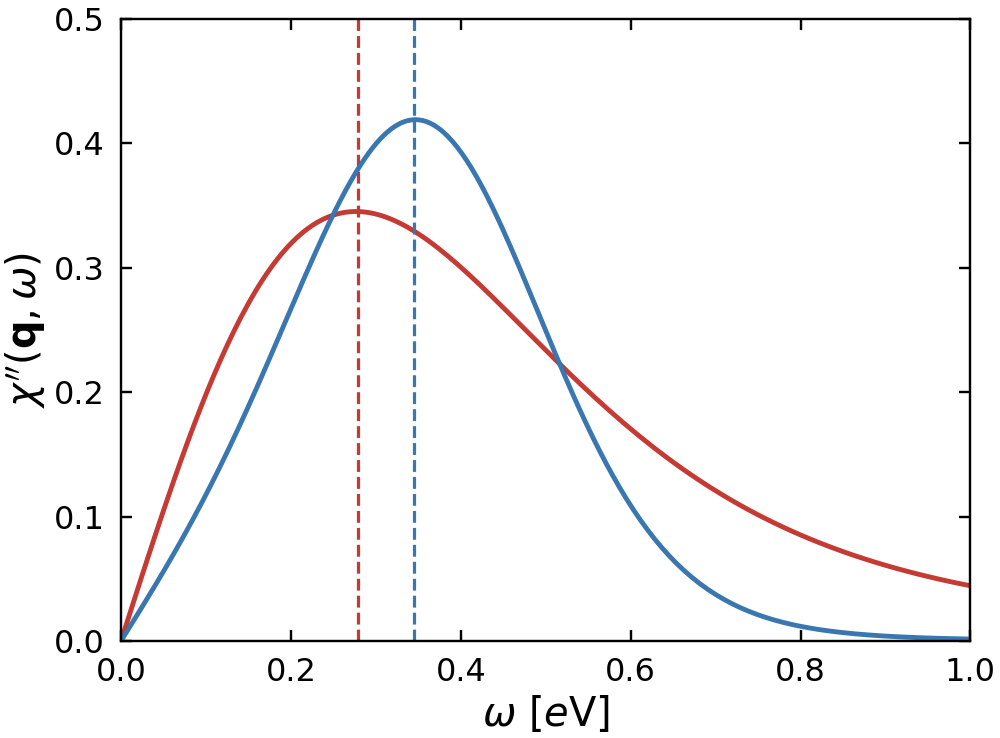}  
  \caption{\label{fig:sqw}Dynamical spin structure factor $\chi''(\textbf{q},\omega)$ for $\textbf{q} = (\pi,0)$. The three-band (red curve) peak position is $2J=0.279~e$V, and that of single-band ($U=8t$, blue curve) is $2J=0.345~e$V.
  }
\end{figure}

For the data we present in the main text, we use $t_{pd} = 1.13~e\text{V}, t_{pp} = 0.49~e\text{V}, U_{dd} = 8.5~e\text{V}, U_{pp} = 4.1~e\text{V}$ and $\Delta_{pd} = 3.24~e\text{V}$ on a 8$\times$8 square cluster for the Emery model, and $t'/t = -0.25$, $U/t = 8$, $t = 0.4~e$V for the single-band Hubbard model of the same size. The magnon peak at $(\pi, 0)$ of $\chi''(\mathbf{q}, \omega)$ of the Emery model and single-band Hubbard model are very close under this set of parameters, as shown in Fig.~\ref{fig:sqw}. The parameters of the Emery model are consistent with Ref.~\cite{3bparams}. The energy scale of the single-band Hubbard model is chosen such that it is consistent with the experimental magnon peak \cite{Hubbard-t}. 

\section{Details of simulation methods}
\begin{figure}[t!]
    \centering \includegraphics[width=1.0\linewidth]{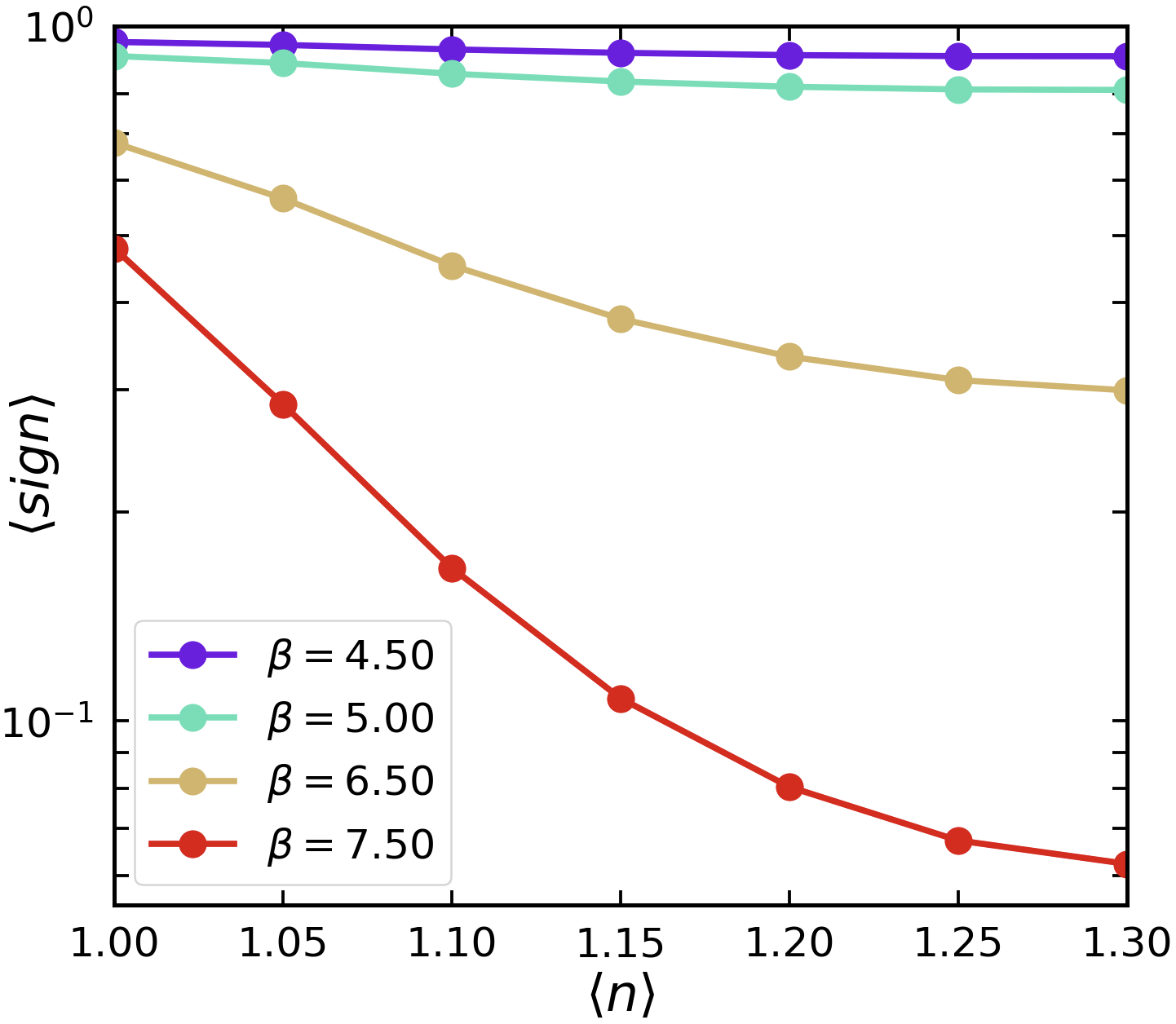}  
  \caption{\label{fig:sign}
  The average fermion sign for three-band $8\times8$ cluster with $U_{dd}=8.5~e$V and $U_{pp}=4.1~e$V.
  }
\end{figure}

In our DQMC simulations for the Emery model, Trotter errors are controlled by setting $\Delta\tau \leq 1/20~e$V$^{-1}$ and $N_{\tau} \geq 80$. The fermion sign at various low temperatures and hole dopings is characterized in Fig.~\ref{fig:sign}. Unless specified otherwise, for all plots in both the supplement and main text, the errorbar indicates $\pm$ one standard deviation from the mean estimated by the jackknife resampling method.

\subsection{A. Optical conductivity}
The frequency-dependent conductivity is obtained from the imaginary-time current-current correlation function 
defined as 
$
\Lambda(\tau)=\langle \mathbf{j}(\tau)\mathbf{j}(0)\rangle 
$, where the current operator at momentum zero
$$\mathbf{j}= i\sum\limits_{i,j,\delta,\delta',\sigma}t_{i\delta,j\delta'}(\mathbf{r}_{i\delta}-\mathbf{r}_{j\delta'})c^\dag_{i,\delta,\sigma}c_{j,\delta',\sigma} + h.c.,$$ with the same definition of $t_{i\delta,j\delta'}$ as in main text. For the three-band model, 
\textbf{$\mathbf{r}_{i\delta}$} labels the coordinates of atoms in the $i$-th unit cell with $\delta$ being the copper $d_{x^2-y^2}$ or oxygen $p_{x,y}$ orbitals. 
Due to C4 symmetry of the square lattice, we consider only the total current in the $x$ direction. The imaginary-time current-current correlation function $\Lambda(\tau)$ measured in DQMC is related to the real-frequency optical conductivity by analytical continuation, 
\begin{equation}
    \Lambda(\tau) = \frac{1}{\pi}\int_{-\infty}^{\infty}\frac{\omega e^{-\tau \omega}}{1 - e^{-\beta\omega}}\sigma(\omega)d\omega
    \label{Eqn:AC}
\end{equation}
or
\begin{equation}
    \Lambda(\tau) = \frac{1}{\pi}\int_{0}^{\infty}\frac{\omega (e^{-\tau \omega} + e^{-(\beta - \tau)\omega})}{1 - e^{-\beta\omega}}\sigma(\omega)d\omega
    \label{Eqn:AC-sym}
\end{equation}
for symmetric bosonic correlators. The optical conductivity is extracted via maximum entropy analytic continuation (MaxEnt) \cite{AC1,AC2}, with error bars at $\omega=0$ determined by bootstrap resampling \cite{Edwin}. 

\begin{figure}[t!]
    \centering \includegraphics[width=1.0\linewidth]{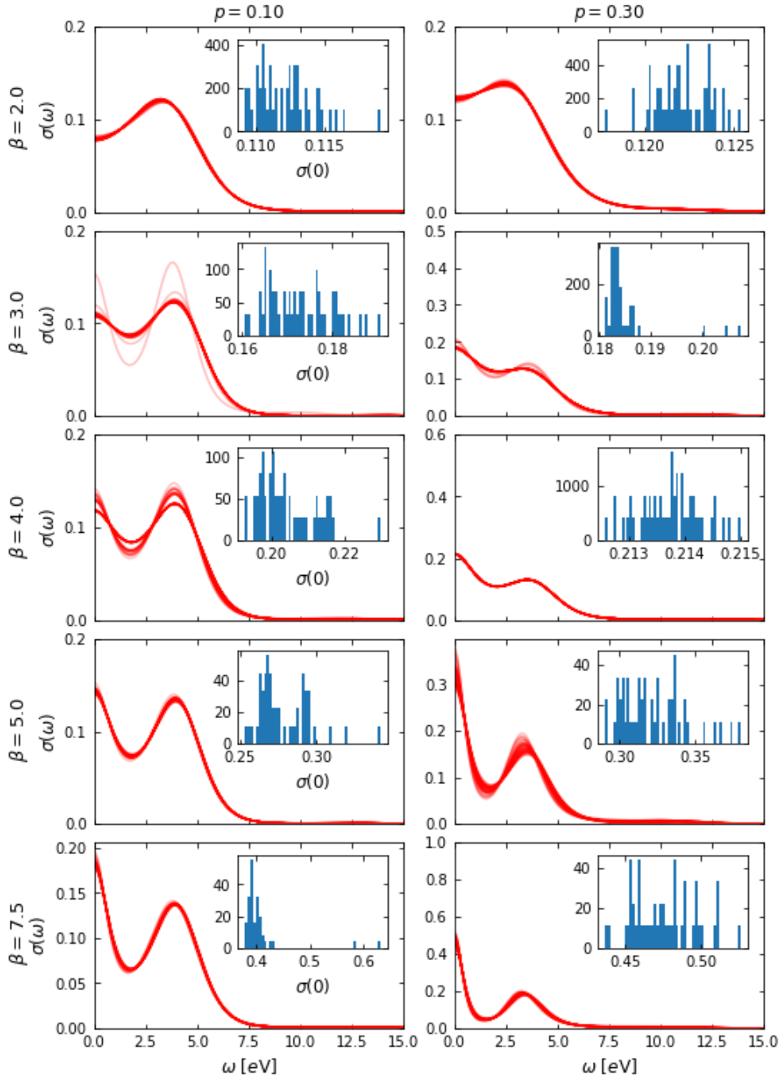}   \caption{\label{fig:histogram}
  Optical conductivity $\sigma(\omega)$ obtained from analytic continuation and bootstrap resampling. Each panel contains 50 resamples. The plot shows results for five characteristic temperatures ranging from $\beta=7.5~e\text{V}^{-1}$ to $\beta=2.0~e\text{V}^{-1}$ at two hole doping values (10\% and 30\%).}
\end{figure}
\begin{figure}[t!]
    \centering
  \includegraphics[width=0.60\linewidth]{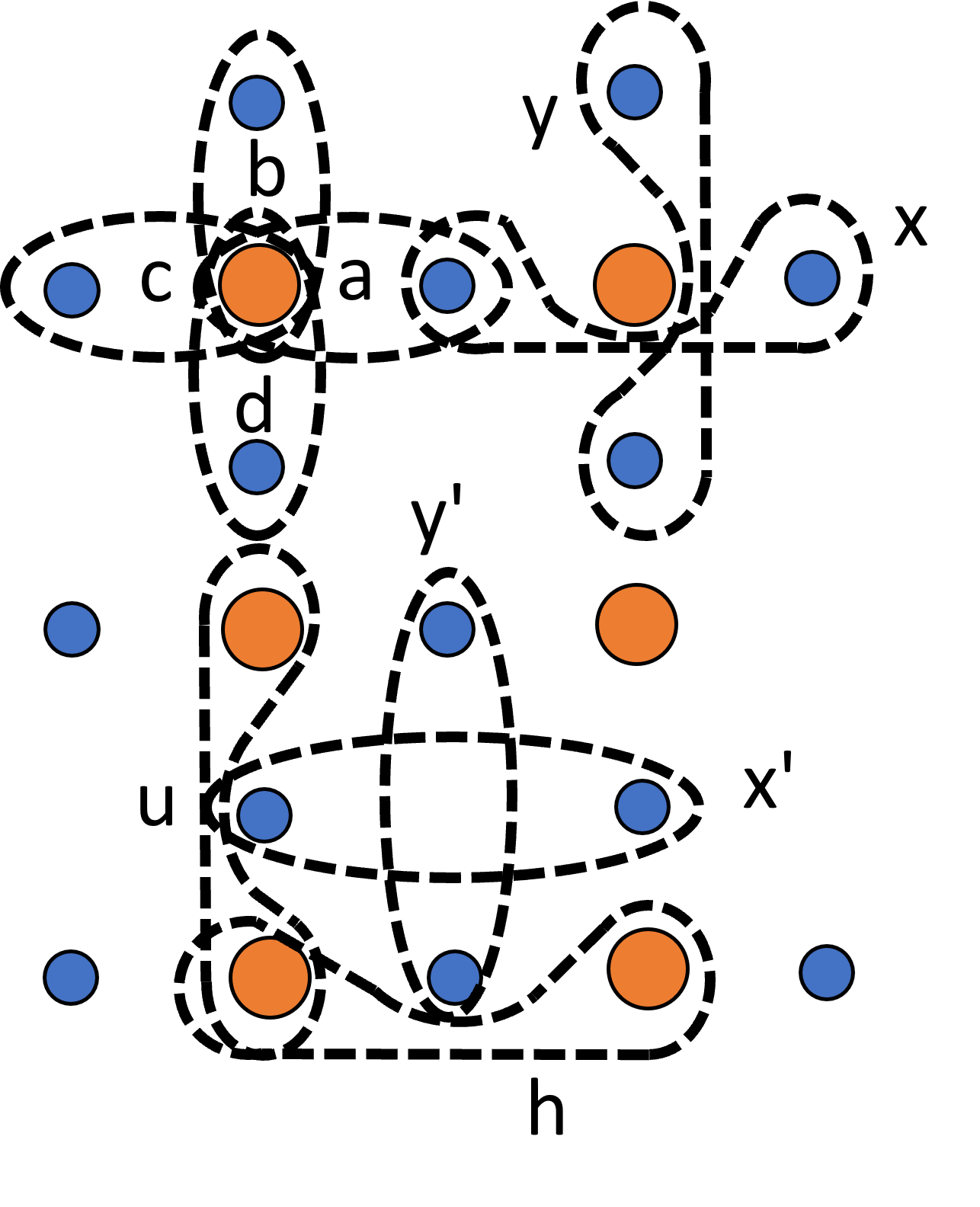}  
  \caption{\label{fig:bonddef} Bond definition on the lattice of three-band model.}
\end{figure}
Figure~\ref{fig:histogram} shows the optical conductivity $\sigma(\omega)$ for the three-band Hubbard model obtained from analytic continuation of bootstrap resampled data. Each small panel contains 50 bootstrap resamples, with insets showing the histograms of $\sigma(\omega=0)$ from resampling.

\subsection{B. Total pair-field susceptibility}
In this section, we elaborate on the definition of the pair-field susceptibility including the Cu and O orbitals. To avoid complications caused by the sign of the hopping parameters of the Emery model, we make the Hamiltonian manifestly four-fold rotational symmetric before we define the pair operators.

The sign convention of the hopping parameters of the Emery model follows Ref.~\cite{Yvonnepaper} and is determined by the sign of the lobes of the $d$ and $p$ orbitals forming a $\sigma$-bond.
By a gauge transformation with wavevector $\Pi = (1,1)\,\pi/a$, the sign of all the hopping parameters can be made the same, and the Emery model is manifestly $C4$ symmetric. Explicitly, the gauge transformation is
\begin{equation}
    \begin{split}
        \widetilde{d}_{i} &= e^{-i\Pi\cdot\vec{R}_{i}} d_{i},\\
        \widetilde{p}_{i}^{x} &= e^{-i\Pi\cdot\vec{R}_{i}} p_{i}^{x},\\
        \widetilde{p}_{i-\hat{y}}^{y} &= e^{-i\Pi\cdot\vec{R}_{i}} p_{i-{\hat{y}}}^{y}
    \end{split}
\end{equation}

\noindent where $d_{i}$, $p_{i}^{x}$, and $p_{i-\hat{y}}^{y}$ have been regrouped into the same unit cell only for the gauge transformation, and $\vec{R}_{i}$ represents the lattice vector pointing from the origin to unit cell $i$. Figure~\ref{fig:bonddef} shows the 10 different singlet operators between nearest-neighbor and next-nearest-neighbor sites in the Emery model at cell $i$. These operators are denoted by $\Delta_{i}^{\mu}$, where the Greek letter $\mu = a, b, c, d, h, u, x, y, x', y'$ indexes the different singlets.
\begin{align}
\begin{split}
    ({\Delta}^{h})_{i} &= \frac{1}{\sqrt{2}} (\widetilde{d}_{i+\mathbf{\hat{x}},\downarrow}\widetilde{d}_{i,\uparrow}  - \widetilde{d}_{i+\mathbf{\hat{x}},\uparrow}\widetilde{d}_{i,\downarrow})\\
    ({\Delta}^{u})_{i} &= \frac{1}{\sqrt{2}} (\widetilde{d}_{i+\mathbf{\hat{y}},\downarrow}\widetilde{d}_{i,\uparrow}  - \widetilde{d}_{i+\mathbf{\hat{y}},\uparrow}\widetilde{d}_{i,\downarrow})\\
    ({\Delta}^{a})_{i} &= \frac{1}{\sqrt{2}}(\widetilde{p}^{x}_{i,\downarrow}\widetilde{d}_{i,\uparrow} -\widetilde{p}^{x}_{i,\uparrow}\widetilde{d}_{i,\downarrow} )\\
    ({\Delta}^{b})_{i} &= \frac{1}{\sqrt{2}}(\widetilde{p}^{y}_{i,\downarrow}\widetilde{d}_{i,\uparrow} - \widetilde{p}^{y}_{i,\uparrow}\widetilde{d}_{i,\downarrow})\\
    ({\Delta}^{c})_{i} &= \frac{1}{\sqrt{2}}(\widetilde{p}^{x}_{i-\mathbf{\hat{x}},\downarrow}\widetilde{d}_{i,\uparrow} - \widetilde{p}^{x}_{i-\mathbf{\hat{x}},\uparrow}\widetilde{d}_{i,\downarrow})\\
    ({\Delta}^{d})_{i} &= \frac{1}{\sqrt{2}}(\widetilde{p}^{y}_{i-\mathbf{\hat{y}},\downarrow}\widetilde{d}_{i,\uparrow} - \widetilde{p}^{y}_{i-\mathbf{\hat{y}},\uparrow}\widetilde{d}_{i,\downarrow})\\
    ({\Delta}^{x})_{i} &= \frac{1}{\sqrt{2}}(\widetilde{p}^{x}_{i,\downarrow}\widetilde{p}^{x}_{i-\hat{x},\uparrow} - \widetilde{p}^{x}_{i,\uparrow}\widetilde{p}^{x}_{i-\hat{x},\downarrow})\\
    ({\Delta}^{y})_{i} &= \frac{1}{\sqrt{2}}(\widetilde{p}^{y}_{i , \downarrow}\widetilde{p}^{y}_{i-\hat{y}, \uparrow} - \widetilde{p}^{y}_{i , \uparrow}\widetilde{p}^{y}_{i-\hat{y}, \downarrow})\\
    ({\Delta}^{x'})_{i} &= \frac{1}{\sqrt{2}}(\widetilde{p}^{y}_{i , \downarrow}\widetilde{p}^{y}_{i-\hat{x}, \uparrow} - \widetilde{p}^{y}_{i , \uparrow}\widetilde{p}^{y}_{i-\hat{x}, \downarrow})\\
    ({\Delta}^{y'})_{i} &= \frac{1}{\sqrt{2}}(\widetilde{p}^{x}_{i,\downarrow}\widetilde{p}^{x}_{i-\hat{y},\uparrow} - \widetilde{p}^{x}_{i,\uparrow}\widetilde{p}^{x}_{i-\hat{y},\downarrow})\\
\end{split}
\end{align}
The definition in the symmetric gauge above translates to the following in the original gauge.
\begin{align}
\begin{split}
    ({\Delta}^{h})_{i} &= -\frac{1}{\sqrt{2}} ({d}_{i+\mathbf{\hat{x}},\downarrow}{d}_{i,\uparrow}  - {d}_{i+\mathbf{\hat{x}},\uparrow}{d}_{i,\downarrow})\\
    ({\Delta}^{u})_{i} &= -\frac{1}{\sqrt{2}} ({d}_{i+\mathbf{\hat{y}},\downarrow}{d}_{i,\uparrow}  - {d}_{i+\mathbf{\hat{y}},\uparrow}{d}_{i,\downarrow})\\
    ({\Delta}^{a})_{i} &= \frac{1}{\sqrt{2}}({p}^{x}_{i,\downarrow}{d}_{i,\uparrow} -{p}^{x}_{i,\uparrow}{d}_{i,\downarrow} )\\
    ({\Delta}^{b})_{i} &= -\frac{1}{\sqrt{2}}({p}^{y}_{i,\downarrow}{d}_{i,\uparrow} - {p}^{y}_{i,\uparrow}{d}_{i,\downarrow})\\
    ({\Delta}^{c})_{i} &= -\frac{1}{\sqrt{2}}({p}^{x}_{i-\mathbf{\hat{x}},\downarrow}{d}_{i,\uparrow} - {p}^{x}_{i-\mathbf{\hat{x}},\uparrow}{d}_{i,\downarrow})\\
    ({\Delta}^{d})_{i} &= \frac{1}{\sqrt{2}}({p}^{y}_{i-\mathbf{\hat{y}},\downarrow}{d}_{i,\uparrow} - {p}^{y}_{i-\mathbf{\hat{y}},\uparrow}{d}_{i,\downarrow})\\
    ({\Delta}^{x})_{i} &= -\frac{1}{\sqrt{2}}({p}^{x}_{i,\downarrow}{p}^{x}_{i-\hat{x},\uparrow} - {p}^{x}_{i,\uparrow}{p}^{y}_{i-\hat{x},\downarrow})\\
    ({\Delta}^{y})_{i} &= -\frac{1}{\sqrt{2}}({p}^{y}_{i , \downarrow}{p}^{y}_{i-\hat{y}, \uparrow} - {p}^{y}_{i , \uparrow}{p}^{y}_{i-\hat{y}, \downarrow})\\
    ({\Delta}^{x'})_{i} &= -\frac{1}{\sqrt{2}}({p}^{y}_{i , \downarrow}{p}^{y}_{i-\hat{x}, \uparrow} - {p}^{y}_{i , \uparrow}{p}^{y}_{i-\hat{x}, \downarrow})\\
    ({\Delta}^{y'})_{i} &= -\frac{1}{\sqrt{2}}({p}^{x}_{i,\downarrow}{p}^{x}_{i-\hat{y},\uparrow} - {p}^{x}_{i,\uparrow}{p}^{x}_{i-\hat{y},\downarrow})\\
\end{split}
\end{align}

\noindent The 10 unique singlet-pairing operators can be partitioned into 4 groups within which the singlet operators are transformed into each other. These 4 groups involve pairing between only Cu orbitals, pairing between Cu and O orbitals, and two components that involve only pairing between O orbitals:
\begin{equation}
\begin{split}
    {\Delta}_{d}^{\text{Cu-Cu}} & := \frac{1}{\sqrt{2}}\sum_{i}\left({\Delta}_{i}^{h}  - {\Delta}_{i}^{u}\right),\\
    {\Delta}_{d}^{\text{Cu-O}} & := \frac{1}{\sqrt{2}}\sum_{i}\left({\Delta}_{i}^{a} - {\Delta}_{i}^{b} + {\Delta}_{i}^{c} -{\Delta}_{i}^{d}\right), \\
    {\Delta}_{d}^{\text{O-O}} & := \frac{1}{\sqrt{2}}\sum_{i}\left({\Delta}_{i}^{x} - {\Delta}_{i}^{y}\right),\\
    {\Delta}_{d}^{\text{O}'\text{-O}'} & := \frac{1}{\sqrt{2}}\sum_{i}\left({\Delta}_{i}^{x'} - {\Delta}_{i}^{y'}\right),
    \label{Eqn:B1gTildeDelta}
\end{split}
\end{equation}
$\Delta_{d}^{\text{O}'-\text{O}'}$ is formed by the pairing of singlets across the Cu orbitals, while $\Delta_{d}^{\text{O-O}}$ is formed by the pairing of singlets across the empty space at the center of the plaquette of the Lieb lattice.
The momentum space representations of the singlet-pairing operators in Eq.~\ref{Eqn:B1gTildeDelta} are given by 
\begin{align}
\begin{split}
    {\Delta}_{d}^{\text{Cu-Cu}} &= \sum_{\mathbf{k}}(\cos(k_{x}) - \cos(k_{y}))\widetilde{d}_{-\mathbf{k},\downarrow}\widetilde{d}_{\mathbf{k},\uparrow}\\
    &= \frac{1}{2}\sum_{\mathbf{k}}(\cos(k_{x}) - \cos(k_{y}))(\widetilde{d}_{-\mathbf{k},\downarrow}\widetilde{d}_{\mathbf{k},\uparrow} - \widetilde{d}_{-\mathbf{k},\uparrow}\widetilde{d}_{\mathbf{k},\downarrow})\\
    &= -\sum_{\mathbf{k}}(\cos(k_{x}) - \cos(k_{y}))d_{-\mathbf{k},\downarrow}d_{\mathbf{k},\uparrow},
    \end{split}
    \end{align}
    \begin{align}
    \begin{split}
    {\Delta}_{d}^{\text{Cu-O}} &= \frac{1}{2}\sum_{\mathbf{k}}[2\cos(k_{x}/2)(\widetilde{p}_{-\mathbf{k},\downarrow}^{x}\widetilde{d}_{\mathbf{k},\uparrow} - \widetilde{p}_{-\mathbf{k},\uparrow}^{x}\widetilde{d}_{\mathbf{k},\downarrow})\\
    &- 2\cos(k_{y}/2)(\widetilde{p}_{-\mathbf{k},\downarrow}^{y}\widetilde{d}_{\mathbf{k},\uparrow} - \widetilde{p}_{-\mathbf{k},\uparrow}^{y}\widetilde{d}_{\mathbf{k},\downarrow})]\\
    &= -\frac{i}{2} \sum_{\mathbf{k}}[2\sin (k_{x}/2)(p^{x}_{-\mathbf{k},\downarrow}d_{\mathbf{k},\uparrow} - p^{x}_{-\mathbf{k},\uparrow}d_{\mathbf{k},\downarrow})\\
    &+ 2\sin (k_{y}/2)(p^{y}_{-\mathbf{k},\downarrow}d_{\mathbf{k},\uparrow} - p^{y}_{-\mathbf{k},\uparrow}d_{\mathbf{k},\downarrow}],
    \end{split}
    \end{align}
    \begin{align}
    \begin{split}
    {\Delta}_{d}^{\text{O-O}} = &\sum_{\mathbf{k}} (\cos(k_{x})\widetilde{p}_{-\mathbf{k},\downarrow}^{x}\widetilde{p}_{\mathbf{k},\uparrow}^{x} - \cos(k_{y})\widetilde{p}_{-\mathbf{k},\downarrow}^{y}\widetilde{p}_{\mathbf{k},\uparrow}^{y})\\
    = &\frac{1}{2}\sum_{\mathbf{k}} [\cos(k_{x})(\widetilde{p}_{-\mathbf{k},\downarrow}^{x}\widetilde{p}_{\mathbf{k},\uparrow}^{x} - \widetilde{p}_{-\mathbf{k},\uparrow}^{x}\widetilde{p}_{\mathbf{k},\downarrow}^{x}) \\&- \cos(k_{y})(\widetilde{p}_{-\mathbf{k},\downarrow}^{y}\widetilde{p}_{\mathbf{k},\uparrow}^{y} - \widetilde{p}_{-\mathbf{k},\uparrow}^{y}\widetilde{p}_{\mathbf{k},\downarrow}^{y})]\\
    = &-\sum_{\mathbf{k}} (\cos(k_{x}){p}_{-\mathbf{k},\downarrow}^{x}{p}_{\mathbf{k},\uparrow}^{x} - \cos(k_{y}){p}_{-\mathbf{k},\downarrow}^{y}{p}_{\mathbf{k},\uparrow}^{y}),
    \end{split}
    \end{align}
    and
    \begin{align}
    \begin{split}
    {\Delta}_{d}^{\text{O}'\text{-O}'} = &\sum_{\mathbf{k}} (\cos(k_{x})\widetilde{p}_{-\mathbf{k},\downarrow}^{y}\widetilde{p}_{\mathbf{k},\uparrow}^{y} - \cos(k_{y})\widetilde{p}_{-\mathbf{k},\downarrow}^{x}\widetilde{p}_{\mathbf{k},\uparrow}^{x})\\
     = &\frac{1}{2}\sum_{\mathbf{k}} [\cos(k_{x})(\widetilde{p}_{-\mathbf{k},\downarrow}^{y}\widetilde{p}_{\mathbf{k},\uparrow}^{y} - \widetilde{p}_{-\mathbf{k},\uparrow}^{y}\widetilde{p}_{\mathbf{k},\downarrow}^{y}) \\&- \cos(k_{y})(\widetilde{p}_{-\mathbf{k}\downarrow}^{x}\widetilde{p}_{\mathbf{k},\uparrow}^{x} - \widetilde{p}_{-\mathbf{k},\uparrow}^{x}\widetilde{p}_{\mathbf{k},\downarrow}^{x})]\\
    = &-\sum_{\mathbf{k}} (\cos(k_{x}){p}_{-\mathbf{k},\downarrow}^{y}{p}_{\mathbf{k},\uparrow}^{y} - \cos(k_{y}){p}_{-\mathbf{k},\downarrow}^{x}{p}_{\mathbf{k},\uparrow}^{x}).
    \end{split}
\end{align}

\begin{figure}
    \centering
    \includegraphics[width=0.92\linewidth]{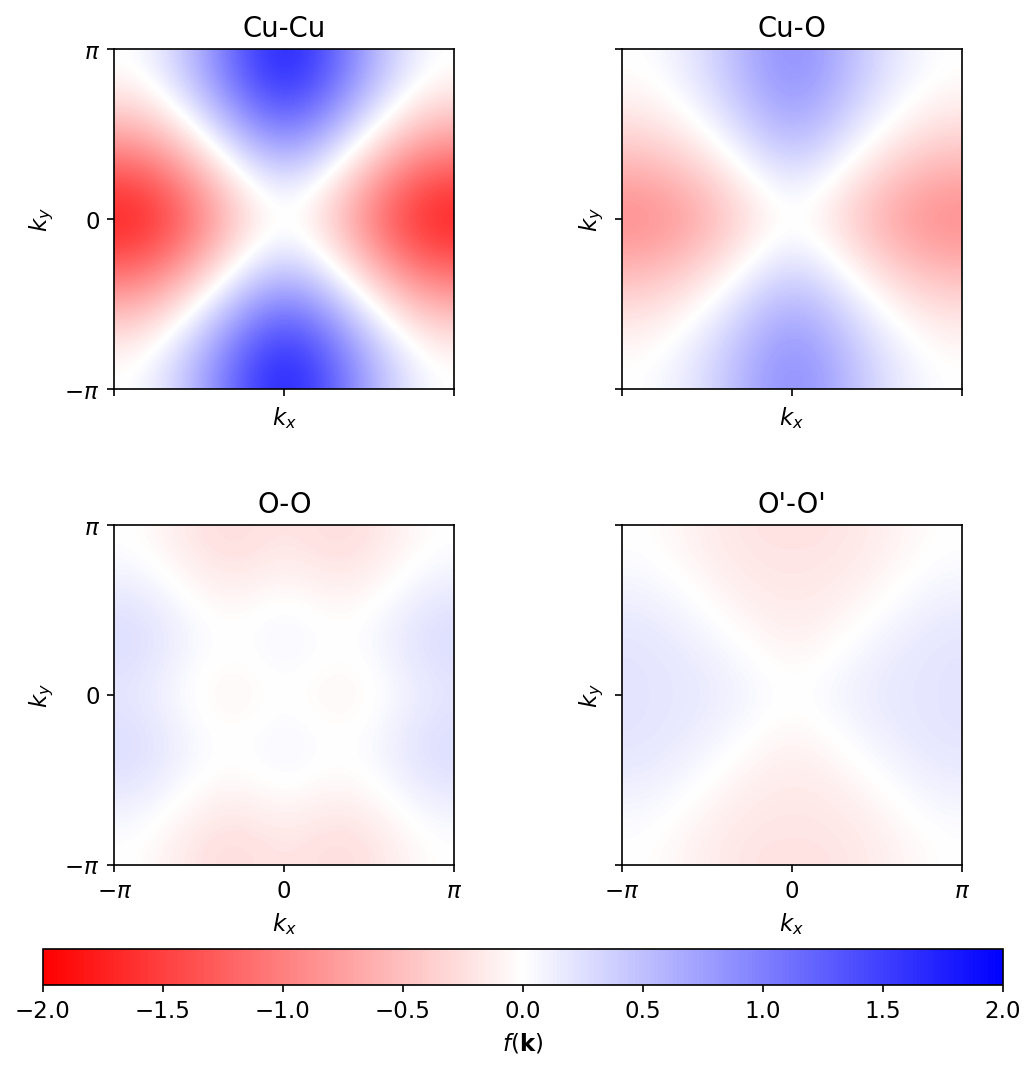}
    \caption{$d$-wave form factor for pair-field susceptibility components $f^{\text{Cu-Cu}}(\mathbf{k}), f^{\text{Cu-O}}(\mathbf{k}), f^{\text{O-O}}(\mathbf{k}), f^{\text{O}'\text{-O}'}(\mathbf{k})$ in non-interacting three-band model.}
    \label{fig:form-factor}
\end{figure}

For the non-interacting three-band model, the effective $d$-wave form-factors with respect to the lowest energy band can be computed numerically and visualized as in Fig.~\ref{fig:form-factor}. The definition of the $d$-wave form factor for each component of the pair-field susceptibility $f^{\text{Cu-Cu}}(\mathbf{k}), f^{\text{Cu-O}}(\mathbf{k}), f^{\text{O-O}}(\mathbf{k}), f^{\text{O}'\text{-O}'}(\mathbf{k})$ is explained in the following. The annihilation operators of each orbital projected to the lowest energy band are
\begin{align*}
    d_{\mathbf{k},\sigma} = \phi_{d,1}(\mathbf{k})c_{\mathbf{k},\sigma}\\
p^{y}_{\mathbf{k},\sigma} = \phi_{y,1}(\mathbf{k})c_{\mathbf{k},\sigma}\\
p^{x}_{\mathbf{k},\sigma} = \phi_{x,1}(\mathbf{k})c_{\mathbf{k},\sigma}
\end{align*}
where $\phi_{d,1}(\mathbf{k}), \phi_{x,1}(\mathbf{k}), \phi_{y,1}(\mathbf{k})$ are the coefficients of the lowest non-interacting energy band, which satisfies $c_{\alpha,\mathbf{k},\sigma} = \sum_{n} \phi_{\alpha, n}(\mathbf{k})c_{n,\mathbf{k},\sigma}$. $\alpha$ denotes the orbital, $n = 1,2,3$ denotes the band index in ascending order in energy (in hole language).  $c_{1,k,\sigma}$ is denoted by $c_{k,\sigma}$ for simplicity. $\phi_{\alpha, n}(k)$ is obtained from diagonalizing the kinetic energy.  The momentum space representation of the $d$-wave superconducting order parameter for the non-interacting three-band model is given by 

\begin{align*}
{\Delta}_{d}^{\text{Cu-Cu}} =  -\sum_{\mathbf{k}}&(\cos(k_{x}) - \cos(k_{y}))\phi_{d,1}(-\mathbf{k})\phi_{d,1}(\mathbf{k})c_{-\mathbf{k},\downarrow}c_{\mathbf{k},\uparrow}\\
= \sum_{\mathbf{k}}& f^{\text{Cu-Cu}}(\mathbf{k}) c_{-\mathbf{k},\downarrow}c_{\mathbf{k},\uparrow},\\
{\Delta}_{d}^{\text{Cu-O}} =  -i \sum_{\mathbf{k}}&[(\sin (k_{x}/2)\phi_{x,1}(-\mathbf{k})+ \sin (k_{y}/2)\phi_{y,1}(-\mathbf{k}))\\&\phi_{d,1}(\mathbf{k})(c_{-\mathbf{k},\downarrow}c_{\mathbf{k},\uparrow} - c_{-\mathbf{k},\uparrow}c_{\mathbf{k},\downarrow})]\\
 = \sum_{\mathbf{k}}& f^{\text{Cu-O}}(\mathbf{k})\frac{1}{2}(c_{-\mathbf{k},\downarrow}c_{\mathbf{k},\uparrow} - c_{-\mathbf{k},\uparrow}c_{\mathbf{k},\downarrow}),\\
{\Delta}_{d}^{\text{O-O}} = -\sum_{\mathbf{k}} &(\cos(k_{x})\phi_{x,1}(-\mathbf{k})\phi_{x,1}(\mathbf{k}) \\&- \cos(k_{y})\phi_{y,1}(-\mathbf{k})\phi_{y,1}(\mathbf{k})){c}_{-\mathbf{k},\downarrow}{c}_{\mathbf{k},\uparrow},\\
 = \sum_{\mathbf{k}}& f^{\text{O-O}}(\mathbf{k}){c}_{-\mathbf{k},\downarrow}{c}_{\mathbf{k},\uparrow}\\
{\Delta}_{d}^{\text{O}'\text{-O}'} = -\sum_{\mathbf{k}} &(\cos(k_{x})\phi_{y,1}(-\mathbf{k})\phi_{y,1}(\mathbf{k}) \\&- \cos(k_{y})\phi_{x,1}(-\mathbf{k})\phi_{x,1}(\mathbf{k})){c}_{-\mathbf{k},\downarrow}{c}_{\mathbf{k},\uparrow}\\
= \sum_{\mathbf{k}}& f^{\text{O}'\text{-O}'}(\mathbf{k}){c}_{-\mathbf{k},\downarrow}{c}_{\mathbf{k},\uparrow}.
\end{align*}

\begin{figure}
    \centering
  \includegraphics[width=1.00\linewidth]{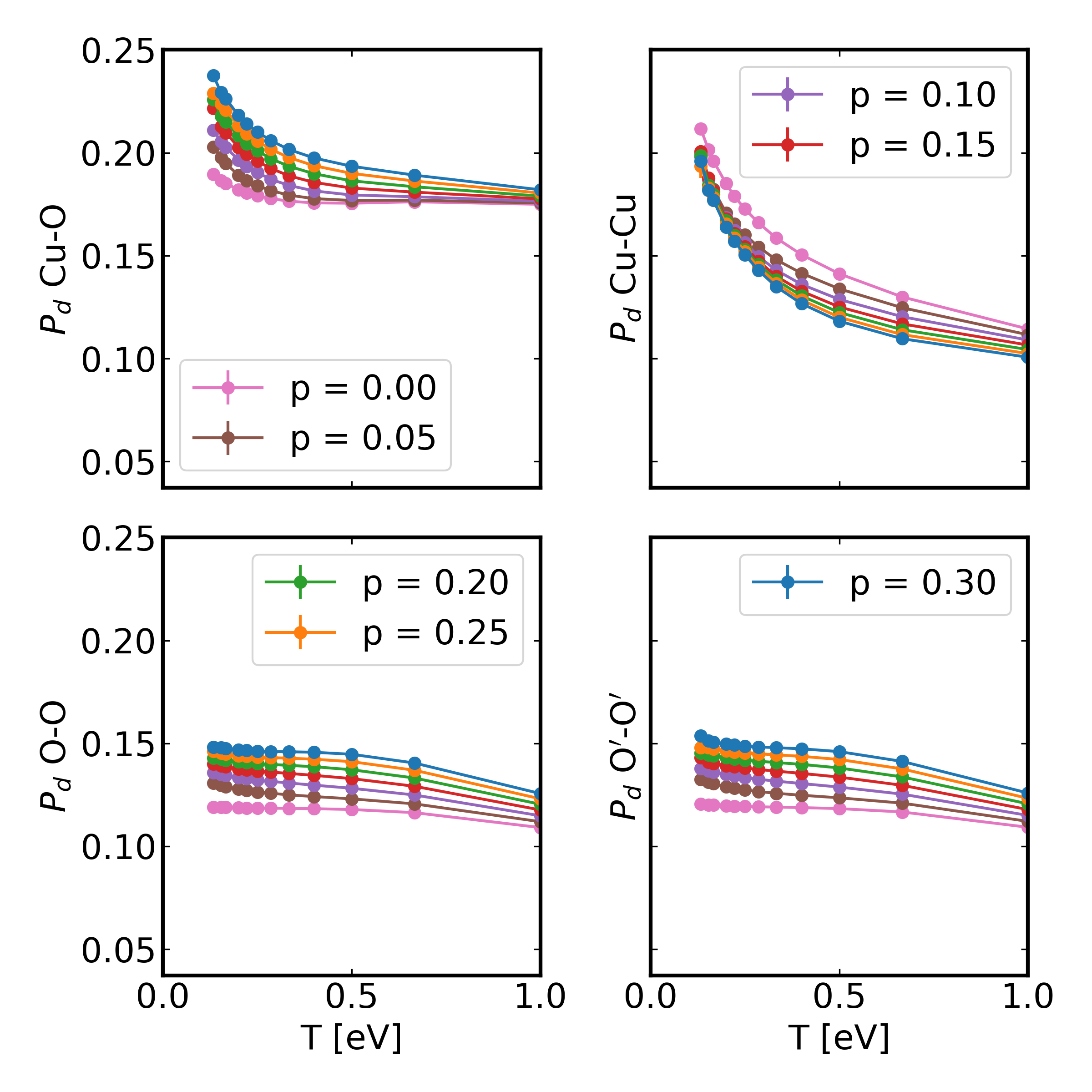}  
  \caption{\label{fig:PdAll} Temperature dependence of the diagonal elements of the $d$-wave pair-field susceptibility matrix  $P_{d}$: $P_{d}^{\text{Cu-O}}$, $P_{d}^{\text{Cu-Cu}}$, $P_{d}^{\text{O-O}}$, and $P_{d}^{\text{O}'\text{-O}'}$.} 
\end{figure}

The $d$-wave pair-field susceptibility matrix $P^{mn}_{d}$ is given by
\begin{align*}
    P_{d}^{mn} &= \langle \Delta_{d}^{m}\Delta^{n\dagger}_{d}\rangle, \text{ where}\\
    m, n &\in \{\text{Cu-Cu}, \text{Cu-O}, \text{O-O},\text{O}'\text{-O}'\}
\end{align*}
The total $d$-wave pair-field susceptibility $P_{\text{total}}$ can be evaluated by the largest eigenvalue of the $4\times 4$ matrix formed by the four components. The temperature dependence of the diagonal elements of the $P_{d}$ matrix are shown in Fig.~\ref{fig:PdAll}. Here, the diagonal elements are plotted because they can be related to the four $d$-wave components defined above.\\

Regarding the pair-field susceptibility as a metric for tendency towards superconductivity, a caveat is that the rate of increasing with decreasing temperature is important, while the magnitude of the pair-field susceptibility is in principle not relevant, although in practice in the absence of phase transition it is hard to distinguish the two factors. 
\begin{figure}
    \centering
    \includegraphics[width=1.00\linewidth]{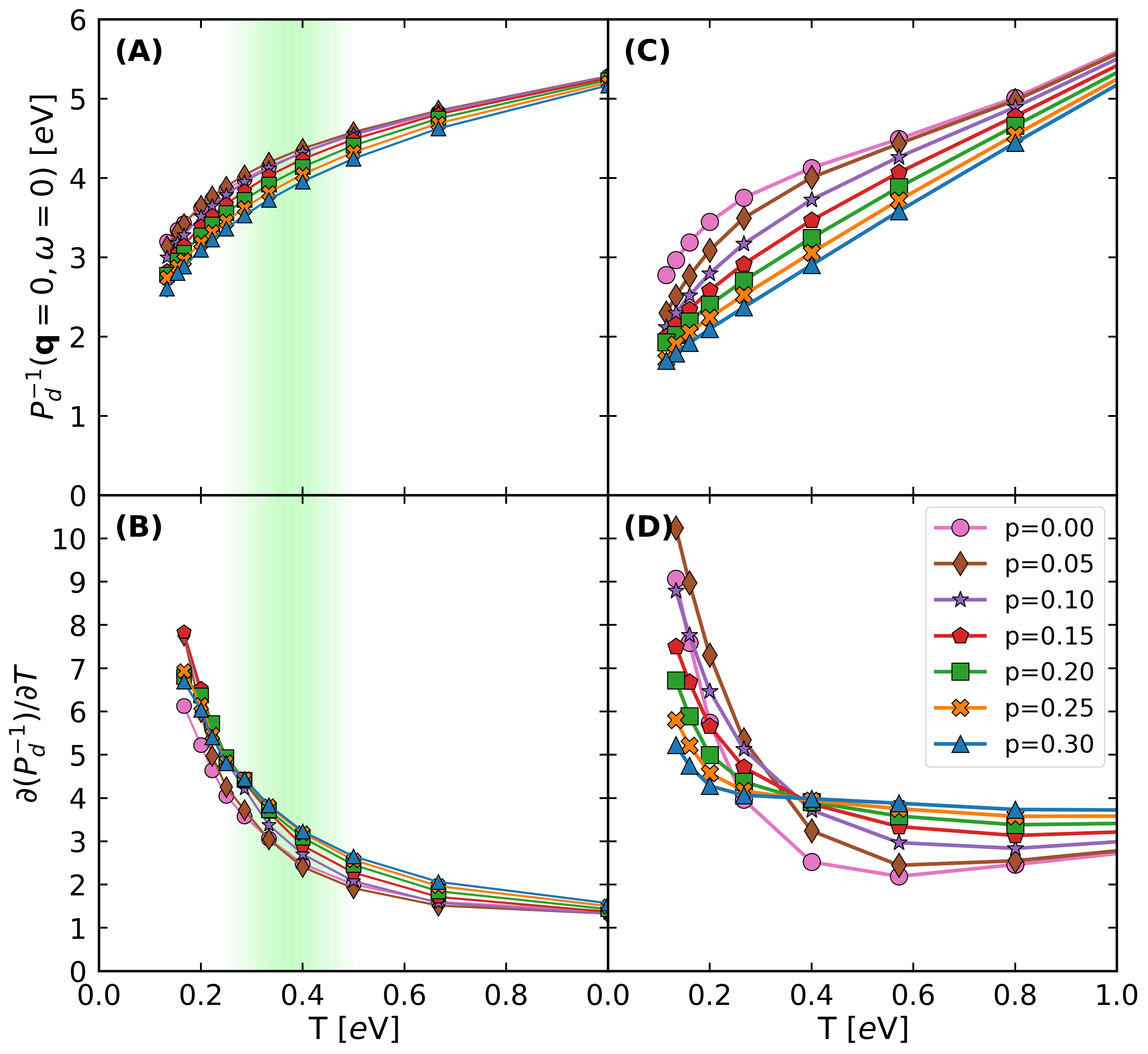}
    \caption{\textbf{Temperature dependent pair-field susceptibility comparison between the three-band Emery (Panels A, B) and single-band Hubbard (Panels C, D) models for different hole doping concentrations $p$.} (A, C) Inverse $d$-wave pair-field susceptibility $P^{-1}_d$ as a function of temperature. (B, D)
Slopes of inverse pair-field susceptibility $\partial(P_d^{-1})/\partial T$ for the data shown in (A, C).}
    \label{fig:SC-nonorm}
\end{figure}

Fig.~\ref{fig:SC-nonorm} shows the temperature derivative of inverse pair-field susceptibility without normalizing against their high temperature values. Clearly at high temperature the inverse pair-field susceptibility decreases faster in single-band Hubbard model, and the values of the derivatives are generally larger in (D) than in (B). However, drawing conclusions based on the magnitude of the first derivative is as bad as drawing conclusions based on the magnitude of the inverse susceptibility, because a constant scaling factor of the temperature dependence should not change the tendency toward superconductivity. Nevertheless, despite that the values in (D) is mostly larger than in (B), (C) is generally lower than in (A), implying a stronger tendency towards superconductivity in single-band Hubbard model above the crossover temperature, which is likely due to the large copper orbital Hubbard interaction, echoing high temperature behavior of the diffusivity. At low temperatures, the three-band Hubbard model has weak doping dependence, while in the single-band Hubbard the temperature derivative at large doping is strongly suppressed.\\

\textit{Uncertainty estimation of the maximal eigenvalue:} 
We denote the pair-field susceptibility matrix as $M$, and its Monte Carlo estimation as $\hat{M}$, which is an estimation of the quantity based on all the data from our Monte Carlo simulation. The expectation value of an estimator is denoted by $\mathbb{E}(\cdot)$, which we assume to be the exact value of the quantity. We denote the eigenvalue with the largest magnitude by $\lambda_{\text{max}}(\cdot)$.

Because the largest eigenvalue of a matrix is a sub-additive norm, we have the following inequalities:
\begin{align*}
|\lambda_{\text{max}}(\hat{M})| &\leq |\lambda_{\text{max}}(\hat{M} - \mathbb{E}(M))| + |\lambda_{\text{max}}(\mathbb{E}(M))|,\\
|\lambda_{\text{max}}(\mathbb{E}(M))| &\leq |\lambda_{\text{max}}(\hat{M})| + |\lambda_{\text{max}}(\mathbb{E}(M) - \hat{M})| ,   
\end{align*}
from which the following bound on the deviation of the estimation from Monte Carlo measurement from expectation value can be derived,
\begin{equation*}
|\lambda_{\text{max}}(\hat{M}) - \lambda_{\text{max}}(\mathbb{E}(M))| \leq |\lambda_{\text{max}}(\hat{M} - \mathbb{E}(M))|,
\end{equation*}
By the Gershgorin circle theorem, the largest eigenvalue of a matrix is bounded by the largest sum of the absolute value of each of the matrix along a row
\begin{equation*}
    |\lambda_\mathrm{max}(\hat{M} - \mathbb{E}(M))| \leq \max_{i} \sum_{j}|\hat{M}_{ij} - \mathbb{E}(M)_{ij}|.
\end{equation*}

$(\hat{M} - \mathbb{E}(M))_{ij}$ is a random variable drawn from a normal distribution with a standard deviation approximated by the estimation from Monte Carlo measurements. We denote one standard deviation of elements of a matrix by $\sigma(\cdot)$. An upper bound of the variance of the largest eigenvalue can be derived using the previous results.
\begin{align*}
    \sigma^{2}(\lambda_{\text{max}}(\hat{M}))
    &= \mathbb{E}\left(\lambda_{\text{max}} (\hat{M}) - \lambda_{\text{max}}(\mathbb{E}(M)\right)^{2}\\
    &\leq \mathbb{E}\left(\lambda_{\text{max}}(\hat{M} - \mathbb{E}(M))\right)^{2}\\
    &\leq \mathbb{E}\left(\max_{i}\sum_{j=1}^{4}|\hat{M}_{ij} - \mathbb{E}(M)_{ij}|\right)^{2}\\
    &\leq \mathbb{E} \max_{i} \left(\sum_{j=1}^{4}|\hat{M}_{ij} - \mathbb{E}\left(M_{ij}\right)|\right)^{2}\\
    &\leq 4 \mathbb{E} \max_{i}\sum_{j=1}^{4}\left(\hat{M}_{ij} - \mathbb{E}(M_{ij})\right)^2\\
    &\leq 4\sum_{i,j=1}^{4}\sigma^{2}(\hat{M}_{ij} - \mathbb{E}(M_{ij}))
\end{align*}
The identity $\sum_{i=1}^{N}|x_{i}|\leq \sqrt{N\sum_{i=1}^{N}(x_{i})^{2}}$ has been used in the fourth inequality. The right-hand side of the last inequality defines the errorbar in the top panel of Fig.~1 in the main text, which as established above, is greater than or equal to one standard deviation of the Monte Carlo estimate of the largest eigenvalue.

\section{Other supplemental data}
\subsection{A. High-temperature DC resistivity}
\begin{figure}[t!]
    \centering  \includegraphics[width=0.8\linewidth]{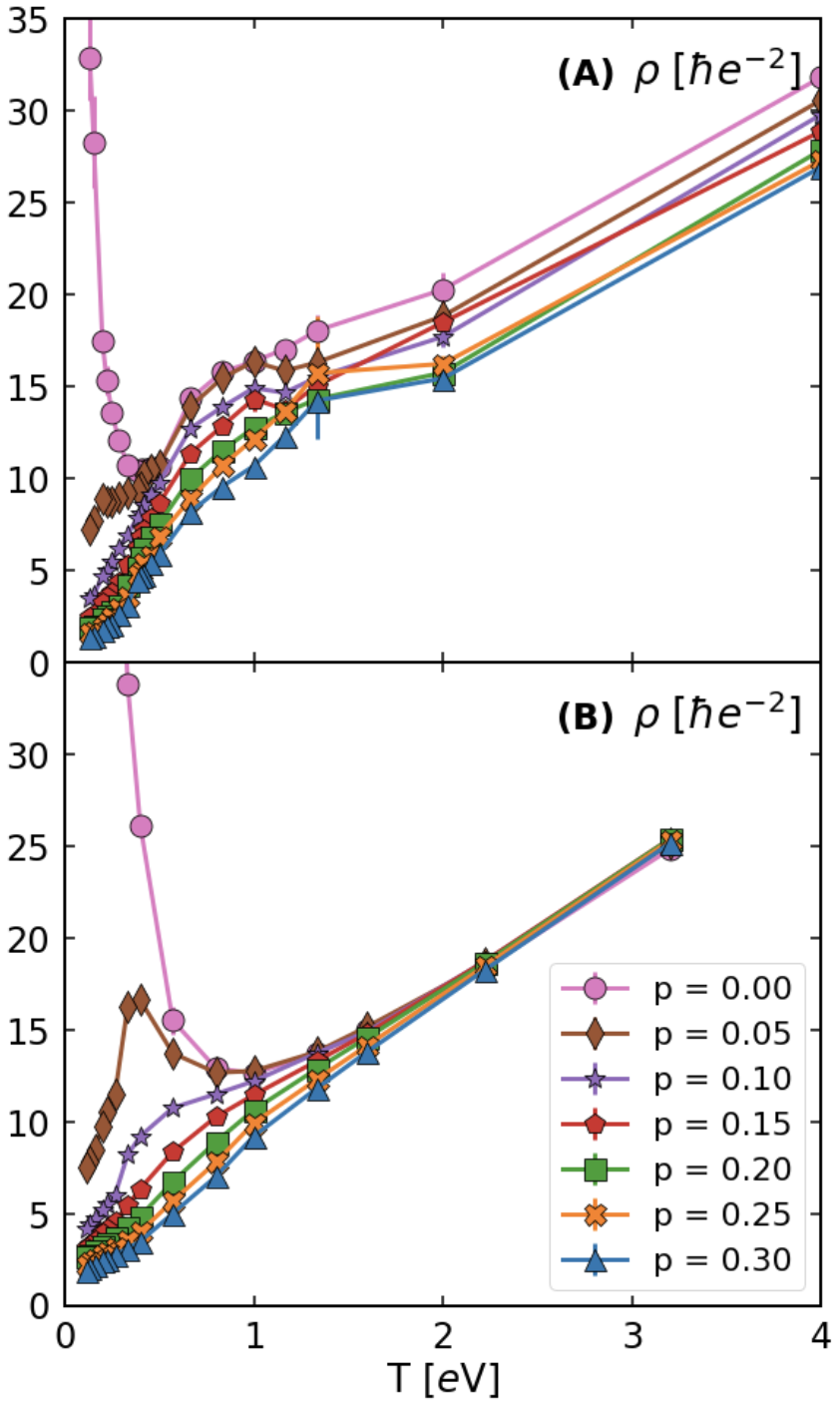} \caption{\label{highTFig4}DC resistivity $\bm{\rho}$ for (A) the three-band Emery model and (B) the single-band Hubbard model over a larger temperature range.}
\end{figure}

The DC resistivities of the three-band and single-band models over a broader temperature range ($0\sim 4~e\text{V}$) are presented in Fig.~\ref{highTFig4}. 

\begin{figure*}[t!]
    \centering  \includegraphics[width=0.8\linewidth]{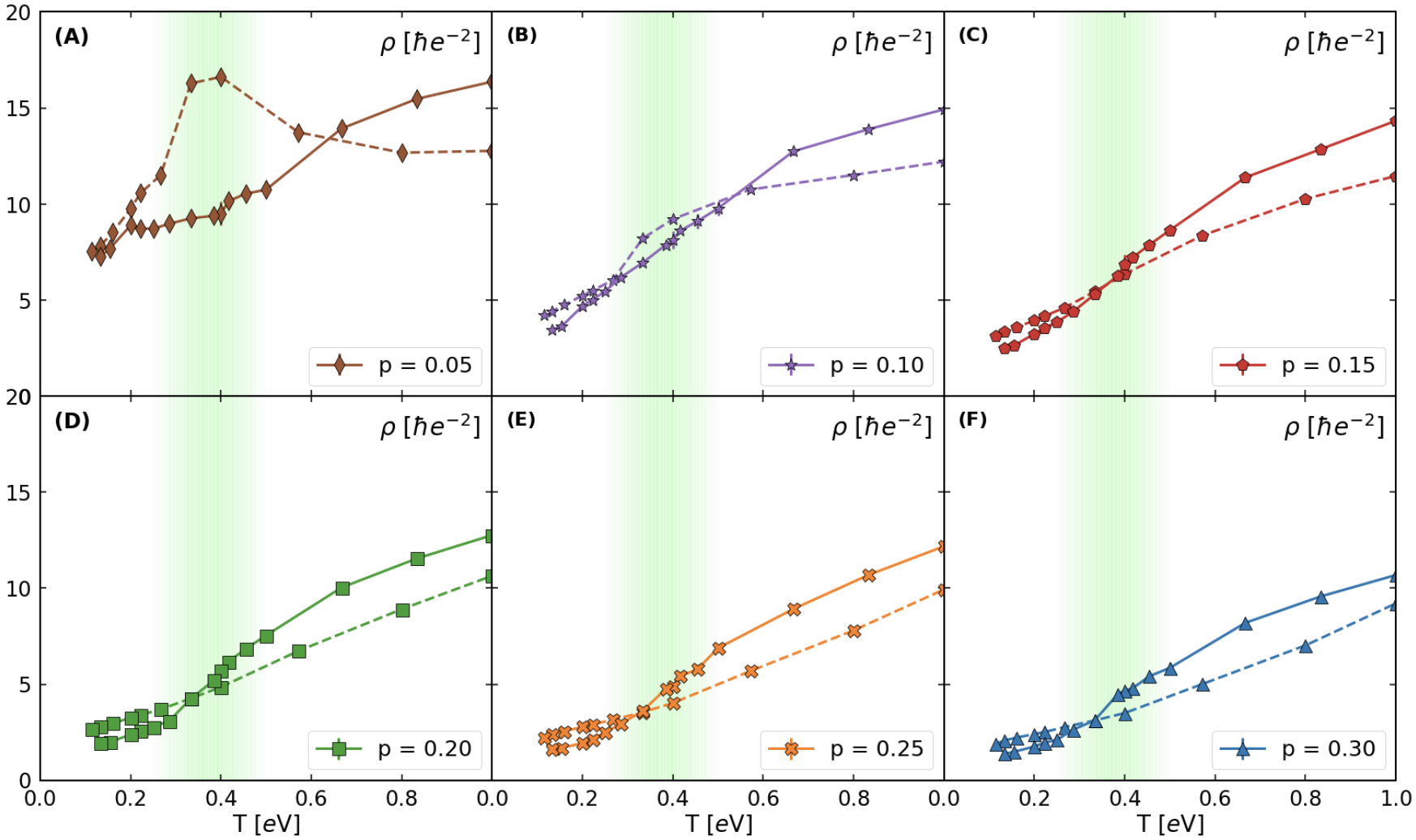} \caption{\label{fig:compare1b3b}\textbf{}DC resistivity $\rho(T)$ for the three-band Hubbard model (solid lines) and the single-band Hubbard model (dashed lines) at six hole dopings p from 5$\%\sim 30\%$. The green-shaded region highlights the inflection (``shoulder'') that appears only in the three-band curves; the single-band response remains essentially linear (constant slope) throughout this range -- except at the lowest dopings ($p=5\%, 10\%)$, where residual Mott-insulator physics produces slight deviations from perfect linearity.}
\end{figure*}
\begin{figure*}[t!]
    \centering  \includegraphics[width=0.925\linewidth]{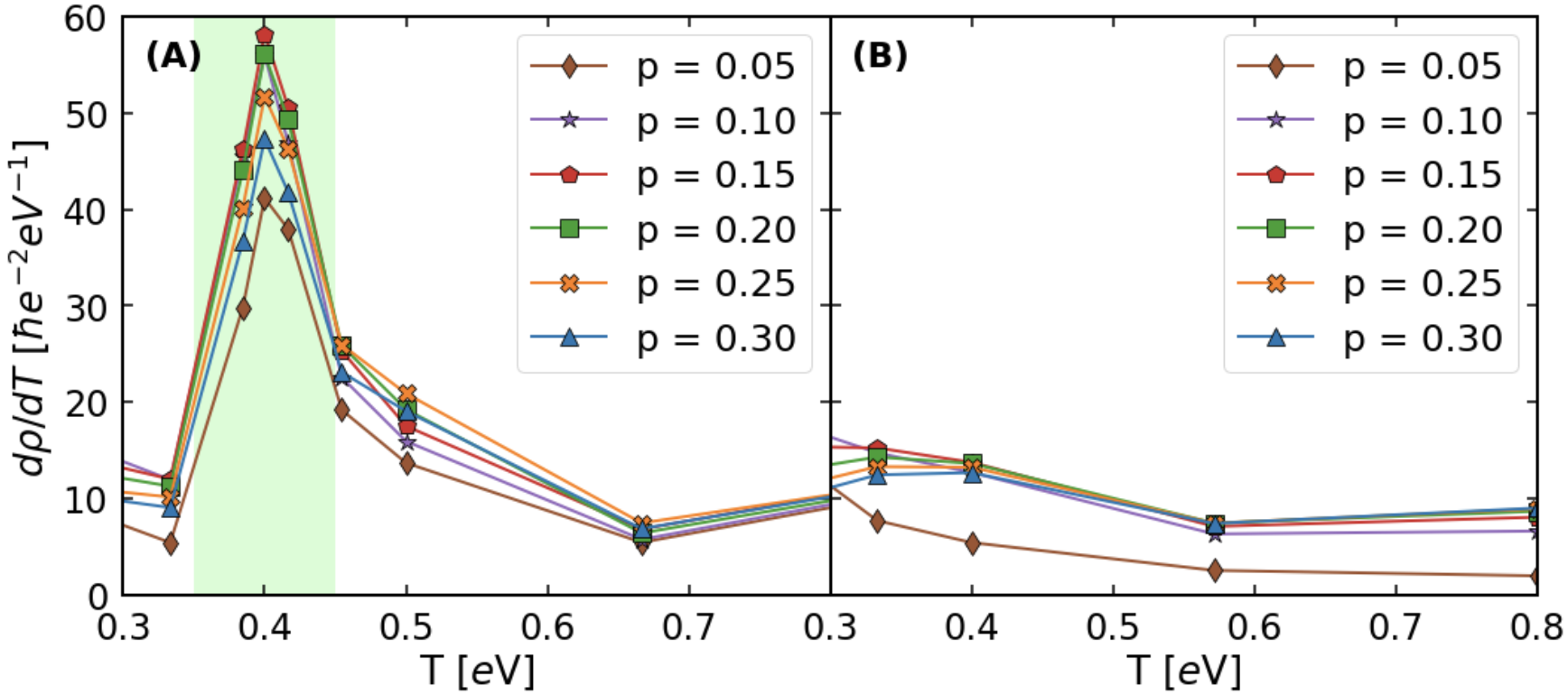} \caption{\label{fig:compare1b3b-2}Temperature derivative of the DC resistivity, $d\rho/dT$, for the three-band model (left) and the single-band model (right). In both cases the derivative was obtained by applying a Savitzky–Golay filter (window=9, polynomial order=1) to the $\rho(T)$ data to smooth out spurious noise and oscillations in the numerical derivative due to analytic continuation. Values are obtained by taking a non-uniform finite difference. In the three-band model each of the resistivity curves for different dopings show a pronounced peak in $d\rho/dT$ at $T \sim 0.4~e$V, which marks the inflection point (``shoulder'') in $\rho$. By contrast, the single-band model exhibits an essentially constant slope over the same temperature range, highlighting the absence of any analogous feature at $0.4~e$V.}
\end{figure*}

\subsection{B. Comparing DC resistivity of single-band and three-band}
Figure~\ref{fig:compare1b3b} shows a doping-by-doping comparison between the single-band and three-band DC resistivity. For $p \sim 0.15$ and above, a crossing occurs at a similar temperature (the ``crossover'' temperature) for all doping, signaling more coherent transport in the three-band model at low temperatures. To further highlight the qualitative differences between the two models, we also compute the temperature derivatives, $d\rho/dT$. As shown in Fig.~\ref{fig:compare1b3b-2}, a pronounced peak in the three-band data occurs at the ``crossover'' temperature while the single-band curves remain essentially flat over the same temperature range.
\begin{figure}[t!]
    \centering  \includegraphics[width=1.0\linewidth]{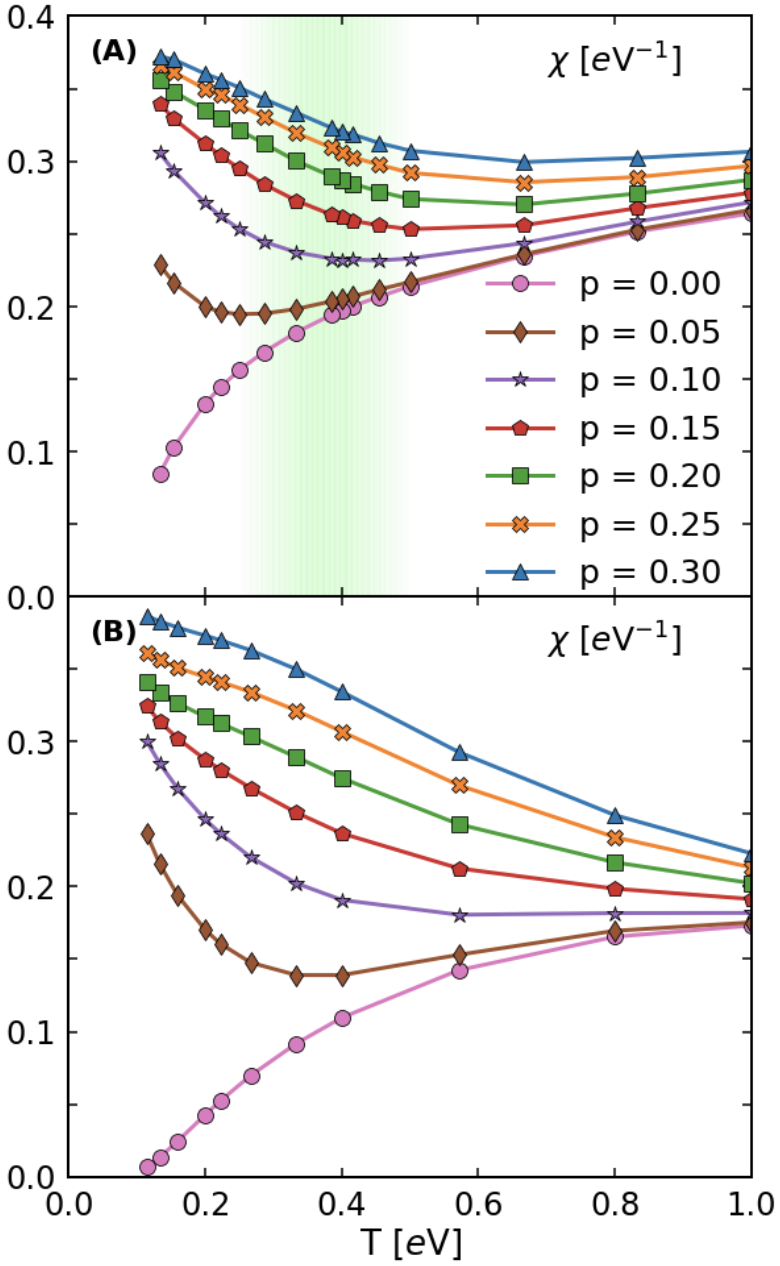} \caption{\label{fig:compare1b3b_chi}A comparison of compressibility $\bm{\chi}$ between (A) the three-band Emery model and (B) the single-band Hubbard model for various values of hole doping.}
\end{figure}

\subsection{C. Comparing compressibility $\chi$ of single-band and three-band}
As mentioned in the main text, via the Nernst-Einstein relation the resistivity can be decomposed into the diffusivity (shown in Fig.~2 of the main text) and the charge compressibility which is plotted in Fig.~\ref{fig:compare1b3b_chi}.

\subsection{D. DC resistivity with \texorpdfstring{$U_{pp}=0$}{Upp=0}}
\begin{figure}[t!]
   \centering  \includegraphics[width=1.0\linewidth]{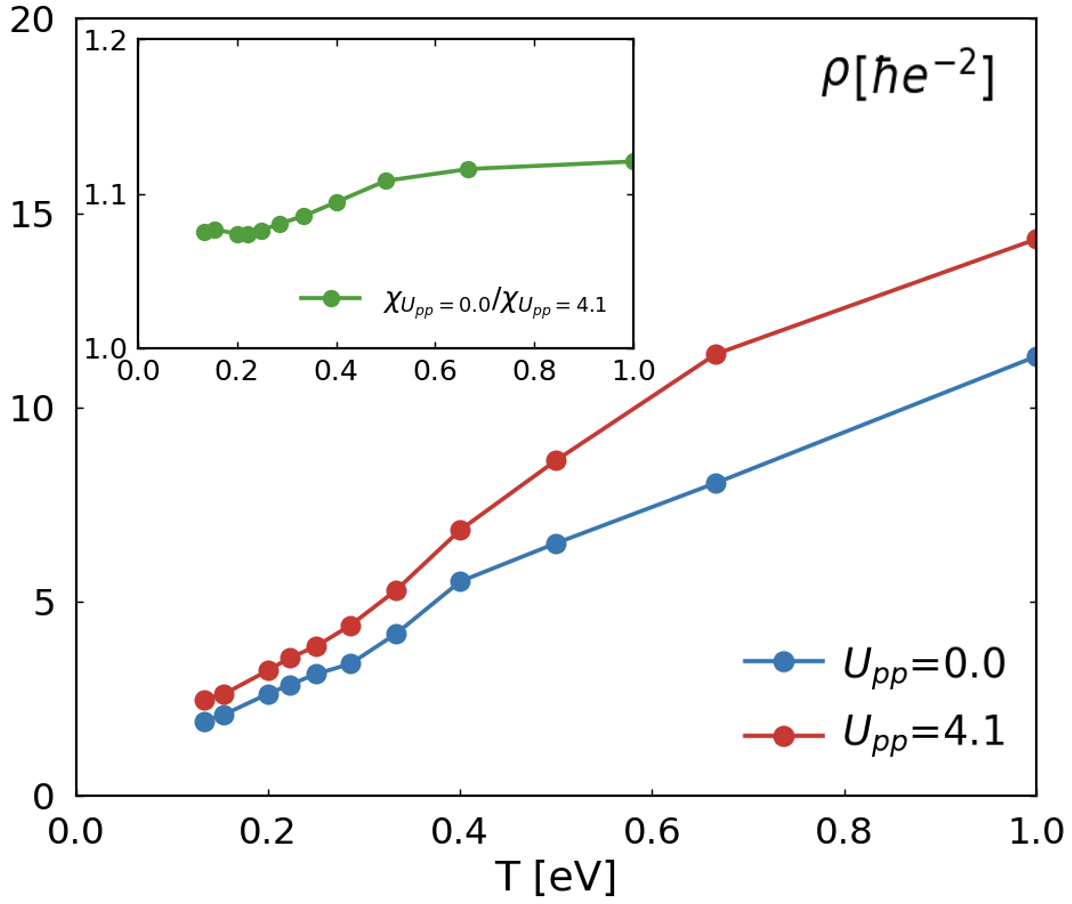} \caption{\label{fig:rhochiD}Comparison between $U_{pp}=4.1~e\text{V}$ and 0 $e\text{V}$ for the DC resistivity and (inset) the ratio of charge compressibility $\chi_{U_{pp}=0.0}/\chi_{U_{pp}=4.1}$ at 15$\%$ hole doping.
   } 
\end{figure}

For the three-band Hubbard model, to investigate the effect of on-site interaction on oxygen orbitals ($U_{pp}$), we compare the DC resistivity with $U_{pp}=4.1~e\text{V}$ and $U_{pp}=0.0~e\text{V}$ together with the ratio of charge compressibility $\chi_{U_{pp}=0.0}/\chi_{U_{pp}=4.1}$ as a function of temperature, as shown in Fig.~\ref{fig:rhochiD}. 

\subsection{E. Cu and O occupation}
\begin{figure}[t!]
    \centering  \includegraphics[width=1.0\linewidth]{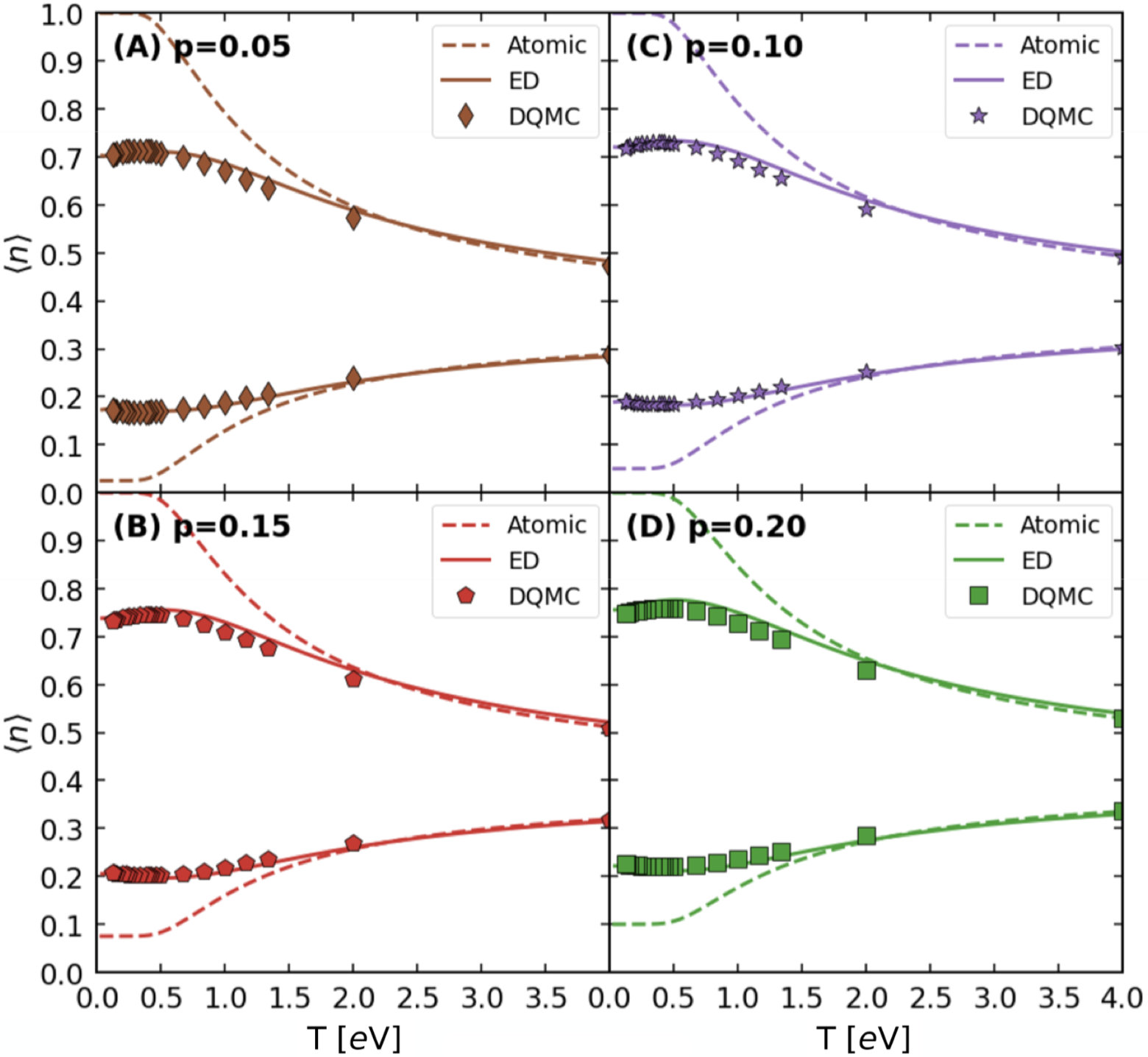}
    \caption{\label{fig:n1}Temperature dependence of Cu and O (Cu occupation is always above O) occupations from DQMC, ED and atomic calculations.
    }
\end{figure}

Figure~\ref{fig:n1} shows the temperature evolution of the hole occupation of Cu and O sites. A maximum (minimum) of Cu (O) occupation is reached at $T \sim 0.4~e$V. Finite-temperature exact diagonalization (ED) is performed on a $2\times2$ CuO$_{2}$ cluster with periodic boundary conditions. The ED results match the DQMC results. At high temperatures, Cu occupation increases with decreasing temperature, and the numerical results overlap with expectations from the atomic limit. This means charge transfer energy plays a dominant role in controlling the orbitally-resolved occupations in the high-temperature limit. Below $T\sim 0.4~e$V, Cu occupation decreases with decreasing temperature, indicating that the kinetic energy is playing a part and promotes hybridization between $d$ and $p$ orbitals. Additional evidence of hybridization can be seen by examining the ED eigenstates in the 5-hole sector (25\% hole doping) where we find that the ground states are spin-1/2 with an average Cu occupation of 0.775, and the first excited states are spin-3/2 with an average Cu occupation of 0.838, 0.3 eV above the ground state. The second excited states have Cu occupations close to the first excited states, but 0.5 eV higher than the first excited state. This indicates that the hybridization forms an effective exchange interaction, which leads to a ground state with a low total spin.
\begin{figure}[t!]
   \centering  \includegraphics[width=0.95\linewidth]{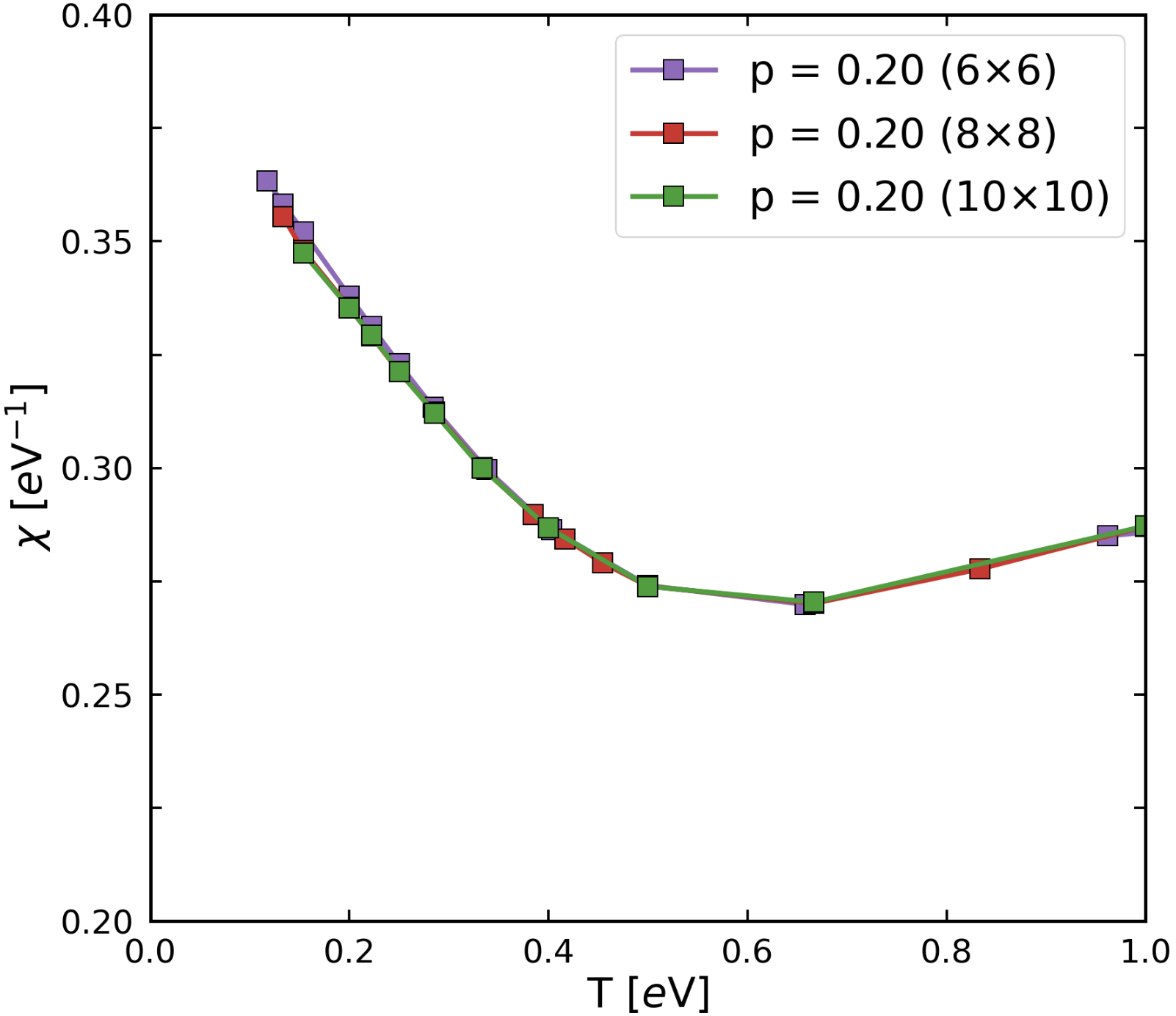} \caption{\label{fig.finite-size}Finite-size analysis of charge compressibility $\chi$ between three different system sizes, $6\times6,~8\times8$ and $10\times10$ at 20$\%$ hole doping.} 
\end{figure}

\subsection{F. Finite-size analysis}
Fig.~\ref{fig.finite-size} shows charge compressibility $\chi$ for three different system sizes: $6\times 6, 8\times 8$ and $10\times 10$ square lattice.

\subsection{G. Magnetic correlation length}
\begin{figure}[t!]
    \centering
    \centering  \includegraphics[width=1.0\linewidth]{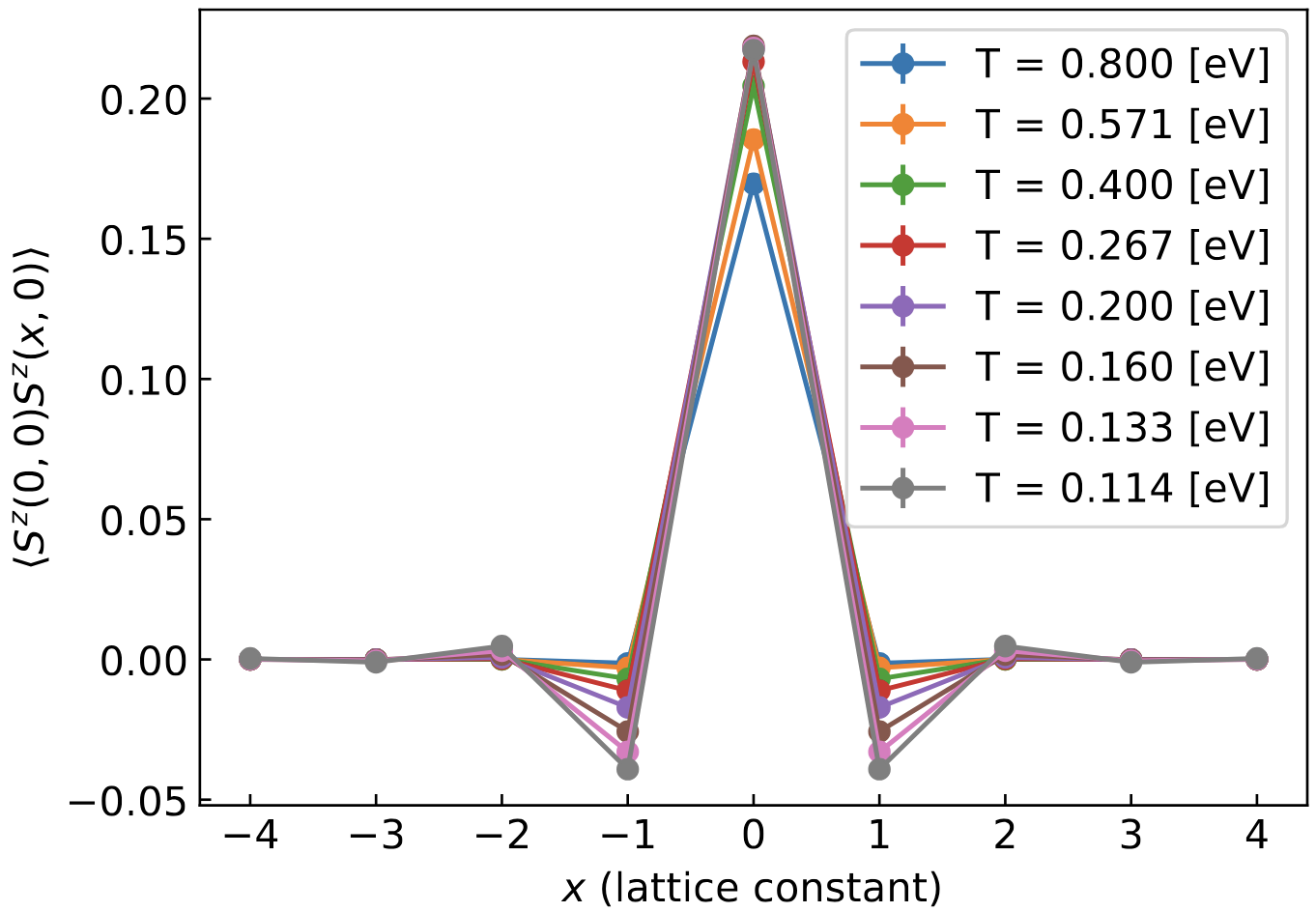}
    \caption{Equal-time spin-spin correlation function in real space along the $x$ axis for the single-band Hubbard model at $p = 0.05$.}
    \label{fig:zz-1b}
\end{figure}

\begin{figure}[t!]
    \centering  \includegraphics[width=1.0\linewidth]{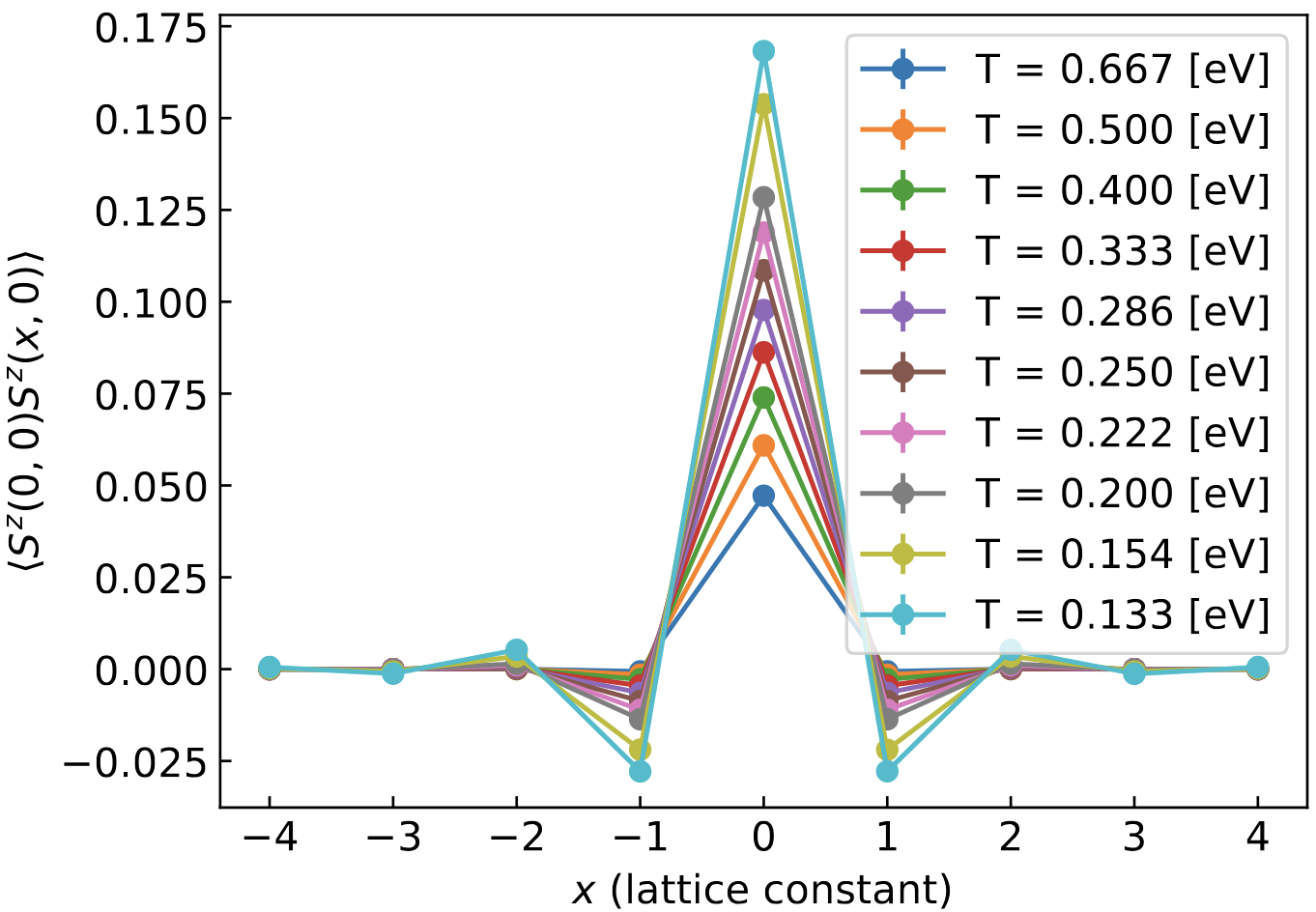}
    \caption{Equal-time spin-spin correlation function in real space along the $x$ axis for the Emery model at $p = 0.05$.}
    \label{fig:zz-1c}
\end{figure}

Figs.~\ref{fig:zz-1b} and~\ref{fig:zz-1c} show that the magnetic correlation lengths are short (less than one lattice constant) and weakly temperature dependent in our temperature range for both the single-band Hubbard model and the three-band Emery model.


\bibliography{bib}